 \documentclass[final,onecolumn,3p,times,sort&compress,12pt]{elsarticle}

\usepackage{amssymb}
\usepackage{hyperref}
\usepackage{epstopdf,xcolor}

\renewcommand*{\today}{October 31, 2016}

\journal{Physica A}

	\usepackage{fancyheadings}
\renewcommand{\thispagestyle}[1]{} 

\begin{document}

\begin{frontmatter}



\title{Hubbard pair cluster in the external fields. Studies of the chemical potential}


\author{T. Balcerzak}
\ead{t\_balcerzak@uni.lodz.pl}
\author{K. Sza{\l}owski\corref{cor1}}
\ead{kszalowski@uni.lodz.pl}
\address{Department of Solid State Physics, Faculty of Physics and Applied Informatics,\\
University of \L\'{o}d\'{z}, ulica Pomorska 149/153, 90-236 \L\'{o}d\'{z}, Poland}

 \cortext[cor1]{Corresponding author. E−mail address:
kszalowski@uni.lodz.pl}

\date{\today}

\begin{abstract}
The chemical potential of the two-site Hubbard cluster (pair) embedded in the external electric and magnetic fields is studied by exact diagonalization of the Hamiltonian. The formalism of the grand canonical ensemble is adopted. The influence of temperature, Hubbard on-site Coulombic energy $U$ and electron concentration on the chemical potential is investigated and illustrated in figures. In particular, a discontinuous behaviour of the chemical potential (or electron concentration) in the ground state is discussed.

\end{abstract}


\end{frontmatter}

\section{Introduction}

The Hubbard model \cite{Anderson, Hubbard, Gutzwiller, Kanamori} has been comprehensively studied  over the last decades \cite{Chen, Ho,  Robaszkiewicz1, Robaszkiewicz2, Hirsch3,Hirsch, Hirsch2,Lieb,Lieb2,Lieb3,Sorella, Pelizzola, Janis, Staudt, Peres, Kent, Zaleski, Schumann,Schumann2, Cisarova,Cencarikova,Galisova,Galisova2,Harris, Silantev, Hasegawa,Hasegawa2,Spalek,Longhi,Juliano,Kozlov, Li,Joura,Alvarez, McKenzie, Fuchs, Rohringer, Kozik, Karchev, Yamada, Claveau, Shastry, Su, Mancini, Tocchio, Dang, Mermin, Nolting, Dombrowsky, Feldner, Szalowski, Barnas, Chao, Yosida, Tasaki, Micnas, Georges, Hirschmeier,Mielke}. Its applicability involves, for example, the description of metal-insulator transition, ferromagnetism of itinerant electrons, studies of high-temperature superconductors, optical lattices and graphene magnetism. In spite of intensive efforts,  so far the exact solution has been obtained only in 1D case \cite{Lieb2, Shastry}, including the solution for the case of Hubbard model in external magnetic field \cite{Su,Mancini}. As far as 2D systems are concerned, the 
Mermin-Wagner theorem about the absence of magnetic ordering for non-zero temperatures is worthy of mention \cite{Mermin, Nolting, Tasaki}. In turn, for the systems 
being in the ground state, the Lieb 
theorems \cite{Lieb}
have been formulated. 

Regarding implications for the magnetism, it has been shown that for the Hubbard parameter $U \to \infty$ (and $U>0$) the so-called  $t-J$ model \cite{Chao} can be derived, which, for half-filling of the band is equivalent to the model of isotropic Heisenberg antiferromagnet \cite{Anderson, Yosida}. In turn, for $U \to \infty$ (and $U<0$) the Hubbard Hamiltonian becomes equivalent to that of XXZ model \cite{Robaszkiewicz1, Micnas}.

Parallel to the exact (or rigorous) results mentioned above, the approximate methods, both analytic and numerical, have been developed with different amount of success. Among them the Quantum Monte Carlo (QMC) methods \cite{Staudt, Kozik} and Dynamical Cluster Approximation (DCA) \cite {Kent} deserve particular attention. 

The possible extensions of the domain where the exact solutions can be found for the model include the zero-dimensional, cluster systems (see for example \cite{Schumann,Schumann2}) as well as clusters embedded in the environment of localized spins \cite{Cisarova,Cencarikova,Galisova,Galisova2}. 
It is worthy to mention that the studies of such geometrically confined, cluster systems are important from the point of view of nanophysics and nanotechnology.
In particular, it has been shown in \cite{Schumann,Schumann2} that for some small number of electrons the analytic solutions can be obtained in addition to the numerical calculations. The magnetic field  has been taken into account in \cite{Schumann}, however, the influence of the external electric field on the cluster properties has not been studied there.

Motivated by such a possibility, we undertake the exact study of the Hubbard simplest cluster, namely the pair (or dimer), embedded simultaneously in the external magnetic and electric fields. We note that such system has already attracted some attention \cite{Harris,Silantev,Hasegawa,Hasegawa2,Spalek,Longhi,Juliano,Kozlov}. Also the effect of the electric field on the properties of the Hubbard model was studied \cite{Li,Joura}. However, the simultaneous influence of both external electric and magnetic field on the Hubbard dimer was not discussed. It can be mentioned that the usefulness of such model can be related to hydrogen molecule \cite{Alvarez} or several layered organic strongly correlated compounds \cite{McKenzie}. We perform analytic diagonalization of the pair Hamiltonian and proceed using the formalism of the grand canonical ensemble, where the mean number of electrons in the system can vary and results from thermodynamic equilibrium conditions. This enables to obtain the grand thermodynamic potential as well as the 
average values of relevant operators. However, 
the statistical-thermodynamic calculations are possible provided the chemical potential is known. Therefore, as the first stage, we found it particularly important to calculate accurately the chemical potential for such an open system, interacting with the environment. 

In this paper we concentrate exclusively on the comprehensive calculations of the chemical potential for the Hubbard pair, especially we study its behaviour in the external magnetic and electric fields, in a wide range of  temperatures. In particular, for low temperature range the quantum changes of the chemical potential are investigated. The influence of Hubbard energy $U$ on the chemical potential is also studied.

In the next section the outline of the theoretical method will be presented. In the successive section the numerical results will be illustrated in figures and their discussion will be given.

\section{Theoretical model}

The Hubbard Hamiltonian for a pair of atoms $(a,b)$ in the external fields  is assumed in the form of:
\begin {eqnarray}
\mathcal{H}_{a,b}&=&-t\sum_{\sigma=\uparrow,\downarrow}\left( c_{a,\sigma}^+c_{b,\sigma}+c_{b,\sigma}^+c_{a,\sigma} \right)+U\left(n_{a,\uparrow}n_{a,\downarrow}+n_{b,\uparrow}n_{b,\downarrow}\right)\nonumber\\
&&-H\left(S_a^z+S_b^z\right) -V\left(n_{a}-n_{b}\right),
\label{eq1}
\end {eqnarray}
where $t>0$ is the hopping integral and $U\ge0$ is on-site Coulomb repulsion energy. The symbol $H=-g\mu_{\rm B}H^z$ introduces an external magnetic field $H^z$, with $g$ being a gyromagnetic factor for the electron and $\mu_{\rm B}$ denoting the Bohr magneton. Moreover, $V=E|e|d/2$ denotes the electrostatic potential of uniform electric field  $E$ oriented along the pair, with $d$ being the interatomic distance, whereas $e$ is the electron charge.  
For the sake of simplicity, we will assume that the hopping integral is not dependent on the external fields. 
In Eq.(\ref{eq1}) $c_{\gamma,\sigma}^+$ and $c_{\gamma,\sigma}$ are the electron creation and annihilation operators, respectively, and $\sigma$ denotes the spin state. The matrix form of the creation and annihilation operators has been derived in Appendix A. Occupation number operators $n_{\gamma}$ for site $\gamma = a,b$, are expressed by these creation and annihilation operators: $n_{\gamma}=\sum_{\sigma}n_{\gamma,\sigma}=\sum_{\sigma}c_{\gamma,\sigma}^+c_{\gamma,\sigma}$, whereas the resulting $z$-components of the spins, $S_{\gamma}^z$, are:  $S_{\gamma}^z=\left(n_{\gamma,\uparrow}-n_{\gamma,\downarrow}\right)/2$.\\

As shown in the Appendix B, the pair Hamiltonian (\ref{eq1}) can be represented by 16 $\times$ 16 matrix.
The exact analytic diagonalization of the Hamiltonian has been performed (see Appendix B). Treating the pair-cluster as an open electronic system, the grand thermodynamic potential $\Omega_{a,b}$ can be obtained from the general formula:
\begin {equation}
\Omega_{a,b}=-k_{\rm B}T \ln \mathcal{Z}_{a,b}=-k_{\rm B}T \ln \{ {\rm Tr}_{a,b} \,\exp \lbrack -\beta \left(\mathcal{H}_{a,b}-\mu\left(n_a+n_b\right)\right)\rbrack \},
\label{eq2}
\end {equation}
where $\mathcal{Z}_{a,b}$ is the grand partition function and $\mu$ is the chemical potential. The knowledge of $\Omega_{a,b}$ enables the self-consistent calculations of all thermodynamic properties.\\

After diagonalization, the thermodynamic mean value of any operator $\hat O$ can be calculated as:
\begin {equation}
\left< \hat O\right>={\rm Tr}_{a,b} \lbrack \hat O \hat \rho_{a,b} \rbrack,
\label{eq3}
\end {equation}
where $\hat \rho_{a,b}$ is the statistical operator for the grand canonical ensemble:
\begin {equation}
\hat \rho_{a,b}=\frac{1}{\mathcal{Z}_{a,b}}\, \exp \lbrack -\beta \left(\mathcal{H}_{a,b}-\mu\left(n_a+n_b\right)\right)\rbrack.
\label{eq4}
\end {equation}
The chemical potential $\mu$ of the electrons fulfils the relationship:
\begin {equation}
\left<n_a\right>+\left<n_b\right>=-\left(\frac{\partial \Omega}{\partial \mu}\right)_{T,H,E}
\label{eq5}
\end {equation}
where $\left<n_a\right>$ and $\left<n_b\right>$ are the thermodynamic mean values of the total occupation number operators for $a$ and $b$ sites, respectively. These averages are parametrized by $\mu$ and can be calculated according to the general formula (\ref{eq3}). The partial derivative in Eq.(\ref{eq5}) is performed at constant temperature $T$ and external fields $H$ and $E$. In order to find $\mu$ we define the parameter $x$ denoting the mean number of electrons per lattice site, namely:
\begin {equation}
x=\left(\left<n_a\right>+\left<n_b\right>\right)/2
\label{eq6}
\end {equation}
Thus, from the relationship  (\ref{eq6}) the chemical potential $\mu$ can be found as a function of $T$, $H$, $E$ and $x$. For a pair of sites, $x$ can take values from 0 to 2, since the system can house up to 4 electrons.\\

The results of numerical calculations  performed in the framework of the above formalism are presented in the next section. 

\section{Numerical results and discussion}

The numerical calculations have been performed on the basis of analytic solution as presented in the theoretical Section. Independently, in order to check analytic formulas, the numerical diagonalization of the pair Hamiltonian has been made using Mathematica software \cite{Wolfram}. The excellent agreement for both methods has been found. The chemical potential has been calculated in a wide range of magnetic and electric fields, for arbitrary temperature, and in a full range of electron concentration $0 \le x \le 2$. For the calculations in the electric field, the external potential $V=V_a=-V_b=E|e|d/2$ has been selected.\\

\begin{figure}[h!]
  \begin{center}
   \includegraphics[scale=0.45]{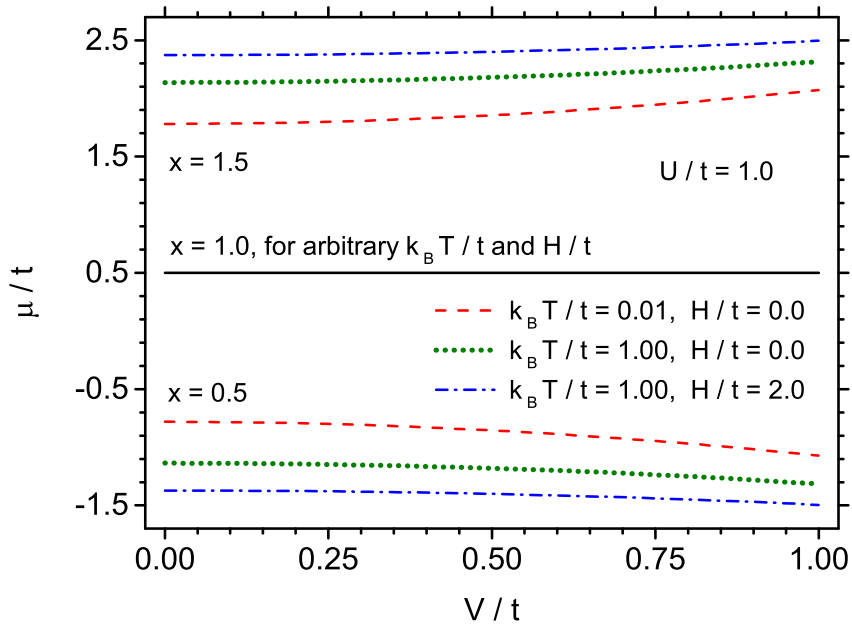}
  \end{center}
   \caption{\label{fig:fig1} Chemical potential $\mu/t$ vs. electric field potential $V/t$ ($V=E|e|d/2$), for $U/t=1$ and different electron concentrations $x$ per atom. Two different temperatures: $k_{\rm B}T/t=0.01$ and $k_{\rm B}T/t=1$, as well as two magnetic fields: $H/t=0$ or $H/t=2$ are chosen.}
\end{figure}

In Fig.~\ref{fig:fig1} the dimensionless chemical potential $\mu/t$ is shown vs. normalized electric field $E|e|d/(2t)$, for two different temperatures $k_{\rm B}T/t$=0.01 and $k_{\rm B}T/t$=1. For higher temperature the effect of magnetic field with the value $H/t$=2 is shown additionally. The curves are plotted for the electron concentrations $x$=0.5, 1 and 1.5. The Hubbard $U$ parameter is set to $U/t=1$. At first it can be noted that for $x=1$ (i.e. for half-filling) the chemical potential is equal to $U/2$ and it does not depend on the temperature and the external fields. This is in agreement with our knowledge about the Hubbard model for infinite system. Moreover, for $x\ne 1$ the results are symmetric with respect to that for $x=1$. For instance, it can be seen that for $x=1.5$, by increasing temperature  the chemical potential also increases, whereas for $x=0.5$ a symmetric decrease of $\mu/t$ is present. The dependence of $\mu/t$ on the electric field is rather weak in this figure and much 
smaller than on the magnetic field. For 
instance, for $k_{\rm B}T/t$=1 a remarkable splitting of two curves: for $H/t$=0 and $H/t$=2 can be noticed. The increasing electric field $E$ makes a slow increase of the chemical potential for $x>1$, whereas for $x<1$ a symmetric slow decrease of $\mu/t$ is observed.\\

\begin{figure}[h!]
  \begin{center}
   \includegraphics[scale=0.45]{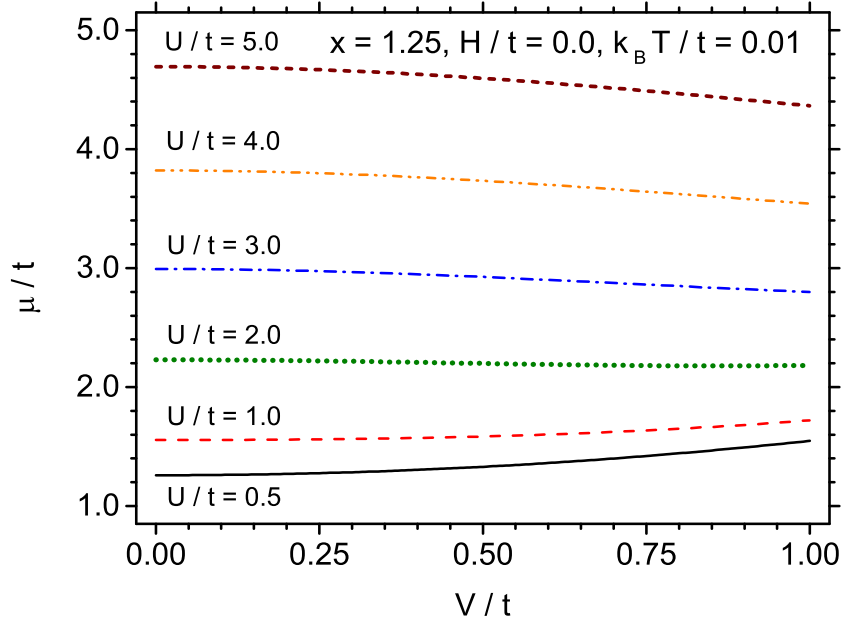}
  \end{center}
   \caption{\label{fig:fig2} Chemical potential $\mu/t$ vs. electric field potential $V/t$ ($V=E|e|d/2$), for electron concentrations $x=1.25$ per atom (weak electron doping) and different $U/t$. Temperature amounts to $k_{\rm B}T/t=0.01$ and the  magnetic field is absent.}
\end{figure}

\begin{figure}[h!]
  \begin{center}
   \includegraphics[scale=0.45]{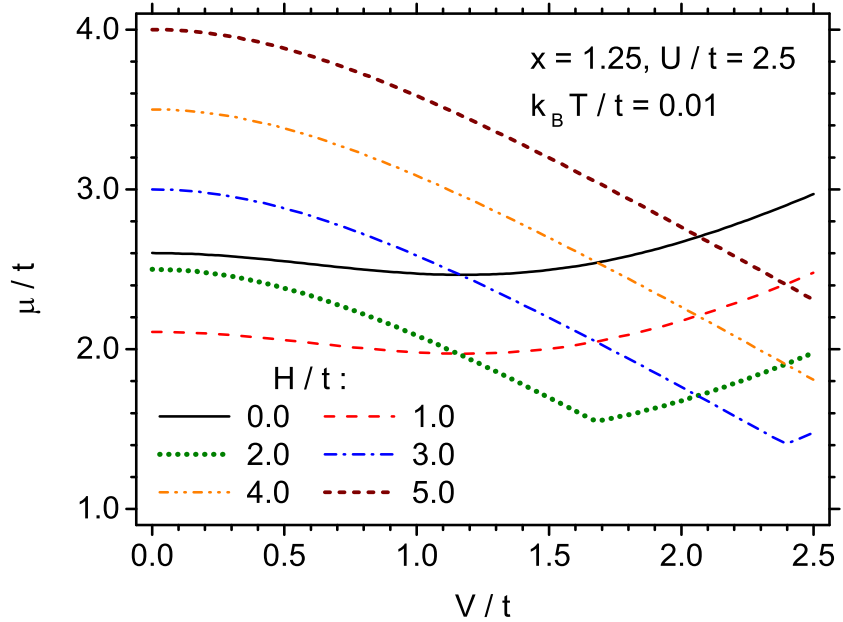}
  \end{center}
   \caption{\label{fig:fig3}Chemical potential $\mu/t$ vs. electric field potential $V/t$ ($V=E|e|d/2$), for electron concentrations $x=1.25$ per atom  and different magnetic field $H/t$. Temperature amounts to $k_{\rm B}T/t=0.01$ and the Hubbard parameter is $U/t=2.5$.}
\end{figure}

\begin{figure}[h!]
  \begin{center}
   \includegraphics[scale=0.45]{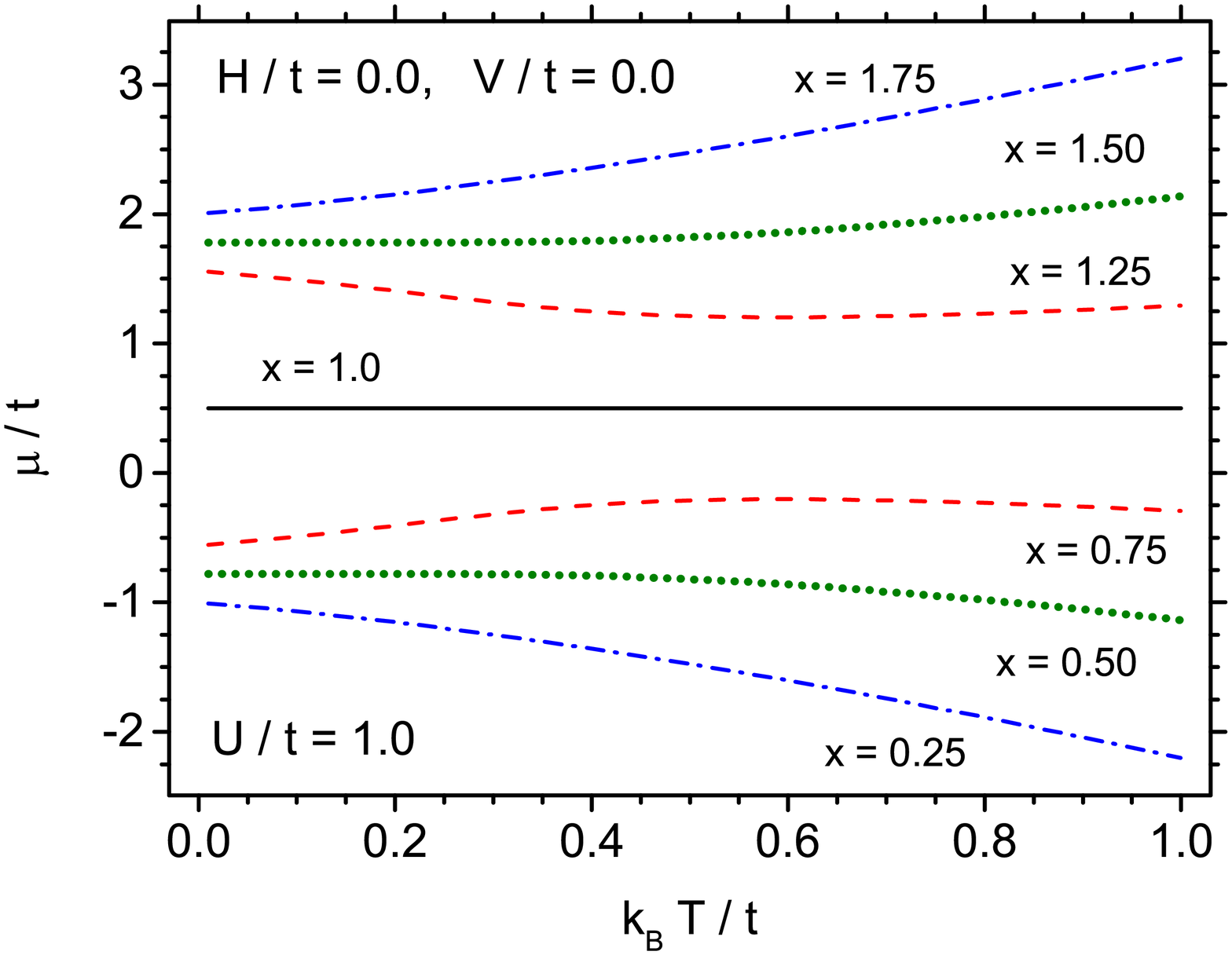}
  \end{center}
   \caption{\label{fig:fig4} Chemical potential $\mu/t$ vs. temperature $k_{\rm B}T/t$, for $U/t=1$ and different electron concentrations $x$ per atom. The magnetic and electric fields are set to zero.}
\end{figure}

The dependence of the chemical potential on the electric field potential, for various Hubbard parameters $U$ is shown in Fig.~\ref{fig:fig2}. In this figure the mean electron concentration is chosen as $x=1.25$, which goes beyond the half-filling, towards small electron doping. The temperature $k_{\rm B}T/t=0.01$ is close to the ground state and the magnetic field is equal to zero. It is seen that the chemical potential $\mu$ strongly depends on the parameter $U$, and increases with an increase of $U$. At the same time, a slowly increasing function of $\mu$ vs. $V$, seen for small $U$, converts into a decreasing one when $U$ increases.\\

A relationship between $\mu$ and the electric field potential $V$ for different magnetic fields is illustrated in Fig.~\ref{fig:fig3}. In this case $U/t=2.5$, whereas the parameter $x$ and temperature are the same as in  Fig.~\ref{fig:fig2}. It is seen that for higher magnetic fields the curves are not monotonous, and have a sharp minimum if the electric field potential $V$ is large enough. On closer inspection of the magnetization, the minima presented for $H/t=2$ and $H/t=3$ are connected with ferrimagnetic phase transitions from saturated phase (with $\left<S_{a}^z + S_{a}^z\right>=0.75$ for low $V$) to the phase with lower total magnetization ($\left<S_{a}^z + S_{a}^z\right>=0.25$), when $V$ increases. The analogous minima occur also for larger fields $H/t$, however, they are not seen in the frame of this figure.\\

In Fig.~\ref{fig:fig4} the chemical potential $\mu/t$ is illustrated vs. dimensionless temperature $k_{\rm B}T/t$  for various electron concentration $x$. The electric and magnetic fields are absent and the $U$ parameter is equal to $U/t=1$. The curves for different $x$ are distributed symmetrically with respect to the curve for $x=1$. For the electron concentration $x$ approaching the value $x=1$ a non-monotonous behaviour of the curves is observed. On the other hand, for $x$ tending to 2 the chemical potential is an increasing function of temperature, whereas for $x$ tending to 0 it monotonously decreases.\\

\begin{figure}[h!]
  \begin{center}
   \includegraphics[scale=0.45]{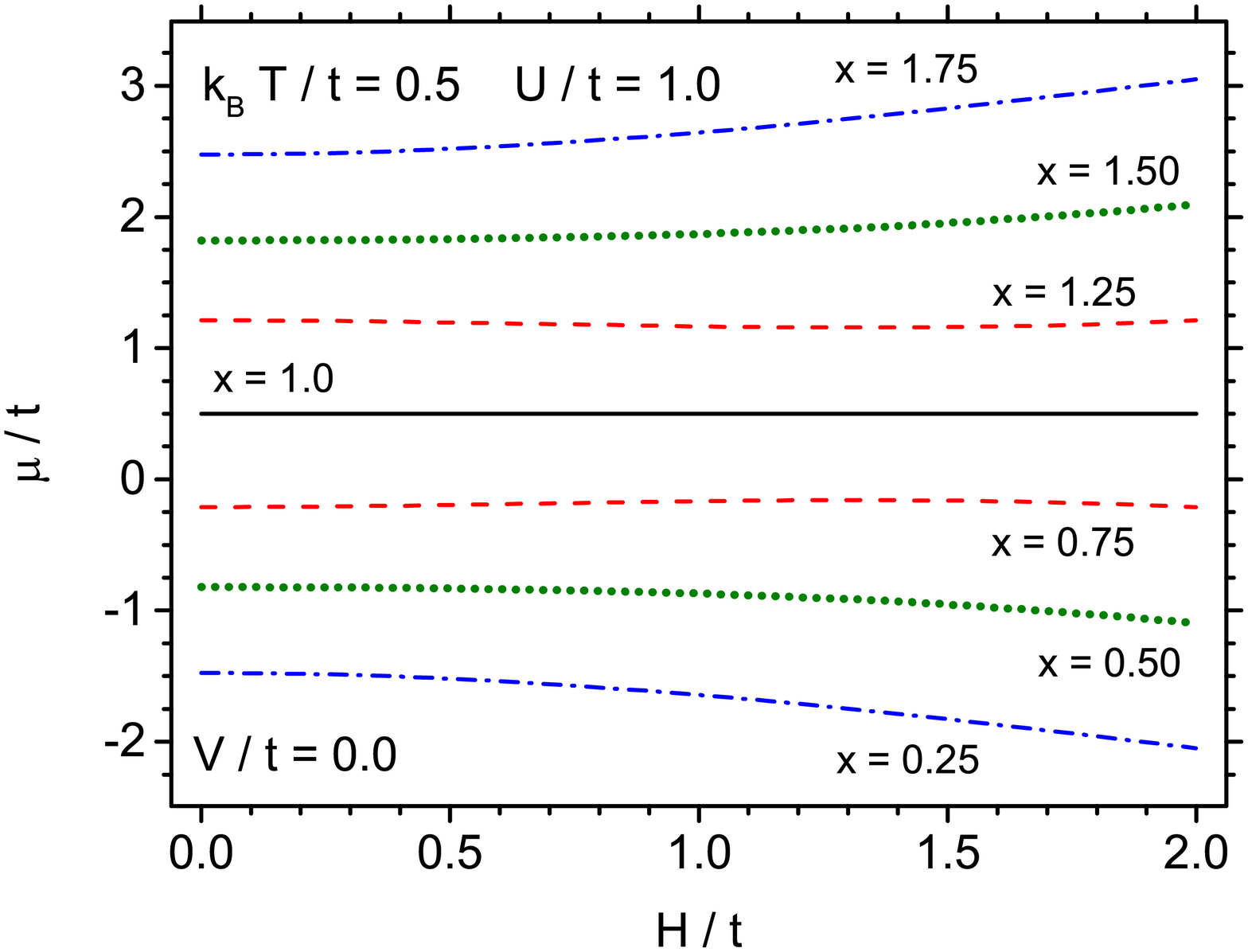}
  \end{center}
   \caption{\label{fig:fig5} Chemical potential $\mu/t$ vs. magnetic field $H/t$, for $U/t=1$ and different electron concentrations $x$ per atom. Temperature is $k_{\rm B}T/t=0.5$ and the electric field is set to zero.}
\end{figure}

In the next figure (Fig.~\ref{fig:fig5}) the chemical potential is presented as a function of the external magnetic field $H/t$, for different electron concentration $x$. The remaining parameters are: $E$=0 and $U/t=1$. Also in this case the symmetry of the curves with respect to the curve for $x=1$ is seen, and non-monotonous behaviour when $x$ approaches the value $x=1$ can be noticed. From the Figs.~\ref{fig:fig4} and \ref{fig:fig5} it is seen that for $x=1$ the chemical potential does not depend on the temperature and magnetic field, in accordance with Fig.~\ref{fig:fig1}.\\

\begin{figure}[h!]
  \begin{center}
   \includegraphics[scale=0.45]{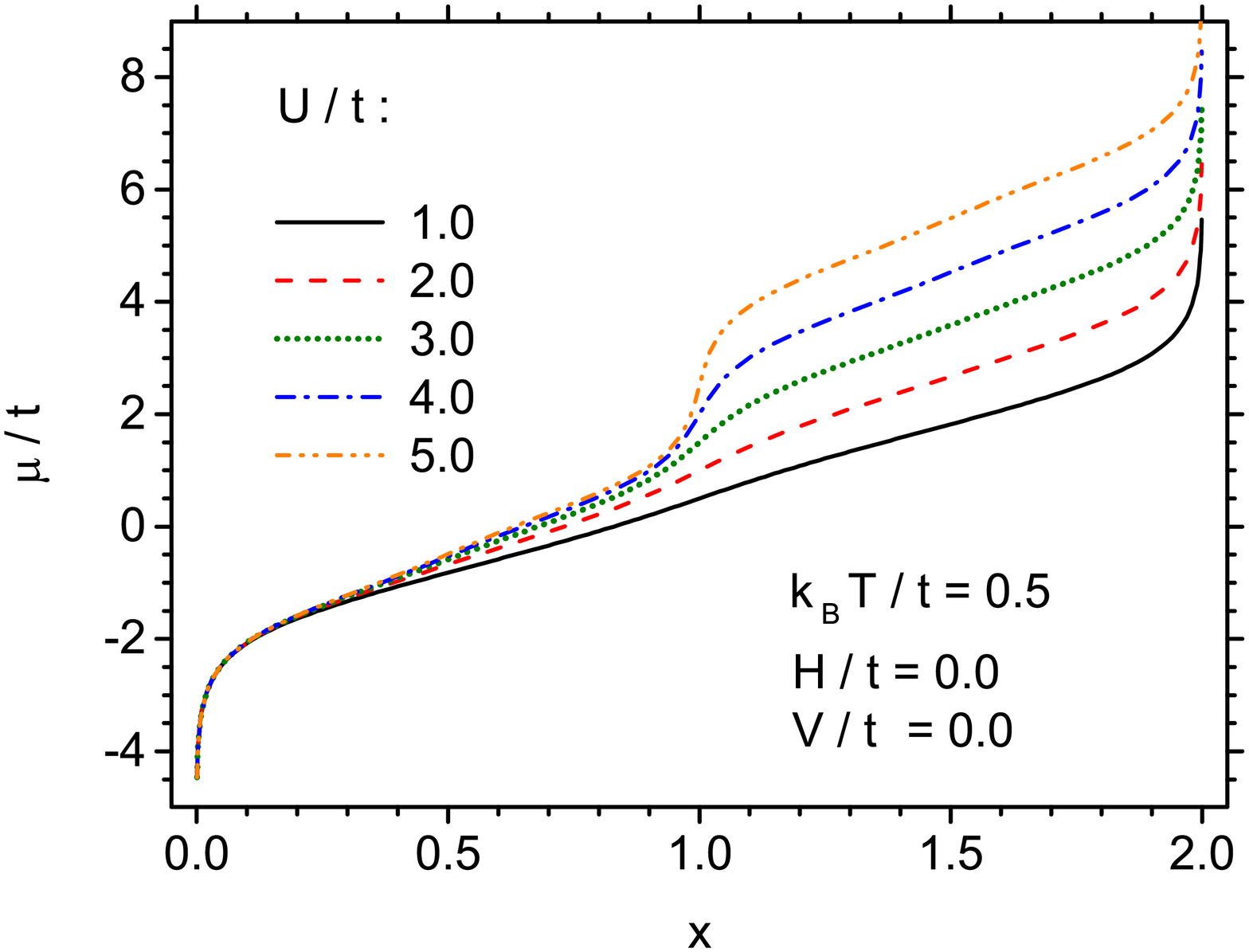}
  \end{center}
   \caption{\label{fig:fig6} Chemical potential $\mu/t$ as a function of electron concentration $x$ per atom, for different Hubbard parameters $U/t$. Temperature is $k_{\rm B}T/t=0.5$ and the magnetic and electric fields are set to zero.}
\end{figure}

The calculation of $\mu/t$ vs. electron concentration $x$ is illustrated in Fig.~\ref{fig:fig6} for different Hubbard energies $U$, at constant temperature $k_{\rm B}T/t$=0.5. 
In Fig.~\ref{fig:fig6} the external fields are equal to zero. All the curves are monotonous vs. $x$, with the strongest dependency for $x \to 0$ and $x \to 2$. The fact that $\partial \mu /\partial x >0$ is important from the thermodynamic point of view, since it proves the chemical stability of the system. For $x=1$ (i.e., for half-filling) the chemical potential fulfils the relation $\mu=U/2$. It can also be noted that for small $x$ the chemical potential is negative and it becomes independent on $U$.\\

\begin{figure}[h!]
  \begin{center}
   \includegraphics[scale=0.45]{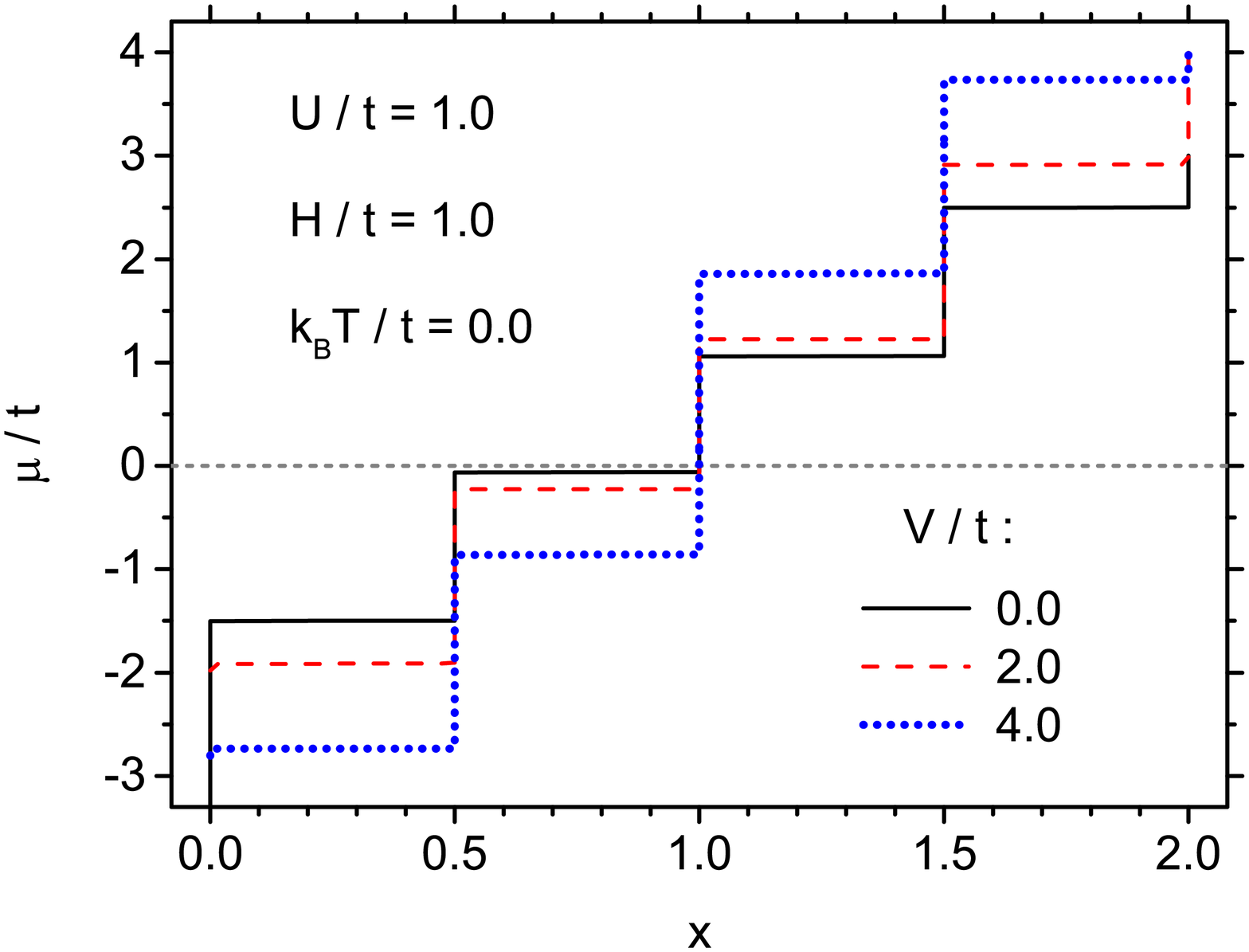}
  \end{center}
   \caption{\label{fig:fig7} Chemical potential $\mu/t$ as a function of electron concentration $x$ per atom, for different  electric field potentials $V/t$. Temperature is $k_{\rm B}T/t=0$, whereas the magnetic field is set to $H/t$=1 and the Hubbard parameter $U/t$=1.}
\end{figure}

In Fig.~\ref{fig:fig7}  the chemical potential $\mu/t$ vs. electron concentration $x$ is presented, for different electric field potentials $V/t$. The magnetic field is also present in this case, with the constant value $H/t$=1, while $U/t$=1. Since the temperature $k_{\rm B}T/t=0$, the chemical potential becomes quantized and is represented by the step-wise function, which shows the steps when the total mean number of electrons in the system, $2x$, amounts to 0, 1, 2, 3 and 4. The influence of electric field potential is evident, manifesting itself by an increasing hight of the steps, whereas the concentrations $x$ at discontinuity points remains the same. The effect of quantization has not been seen in the previous figure (Fig.~\ref{fig:fig6}), because the temperature was too high there.\\

\begin{figure}[h!]
  \begin{center}
   \includegraphics[scale=0.45]{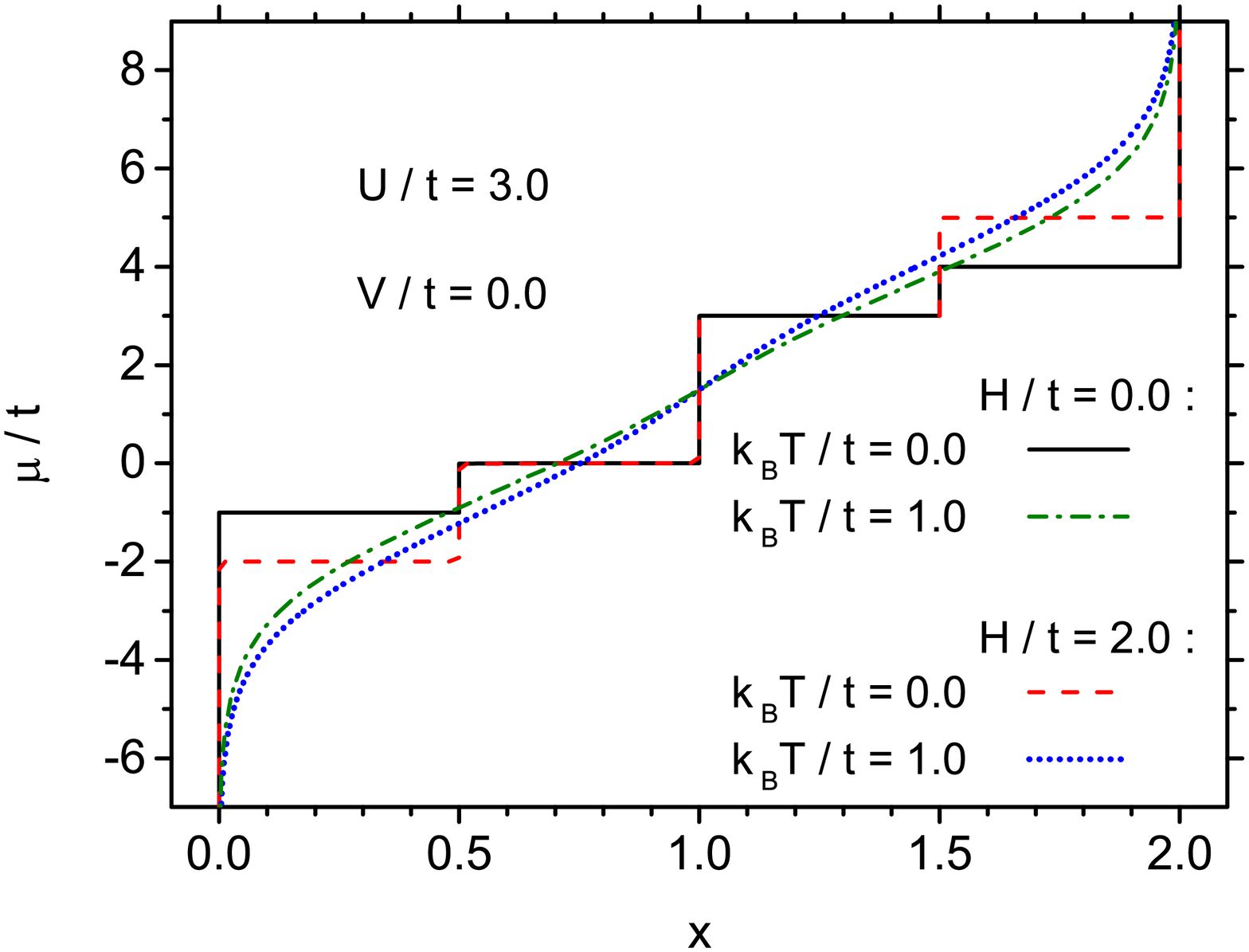}
  \end{center}
   \caption{\label{fig:fig8} Chemical potential $\mu/t$ vs. electron concentrations $x$ per atom, for $U/t=3$ and two different temperatures: $k_{\rm B}T/t=0$ and $k_{\rm B}T/t=1$. Two magnetic fields are chosen: $H/t=0$ and $H/t=2$, whereas the electric field is set to zero.}
\end{figure}

In order to demonstrate the temperature effect on  quantization, the dependency of $\mu/t$ on the electron concentration $x$ is plotted in Fig.~\ref{fig:fig8} for two very different temperatures: $k_{\rm B}T/t$=0 (i.e., in the ground state) and $k_{\rm B}T/t$=1. In addition,  absence of the magnetic field, or its  presence with the value $H/t=2$, is taken into account, whereas $E=0$ and  $U/t$=3. Again, it is seen that in the ground state the chemical potential changes discontinuously when the concentrations take the values: $x=$0, 0.5, 1, 1.5 and 2. As expected, such a quantum behaviour is not present in the high temperature region, where the strong thermal fluctuations  are present. In the ground state, the influence of the magnetic field is seen only in the regions $0 \le x \le 0.5$ and $1.5 \le x \le 2$, whereas no effect on $\mu$ in the region of $0.5 < x < 1.5$ is noticed. On the other hand, for relatively large temperature, $k_{\rm B}T/t$=1, the magnetic field 
influences the chemical potential practically in the whole region of $x$, with the 
exclusion of $x=1$ where $\mu/t=1.5$.\\

\begin{figure}[h!]
  \begin{center}
   \includegraphics[scale=0.45]{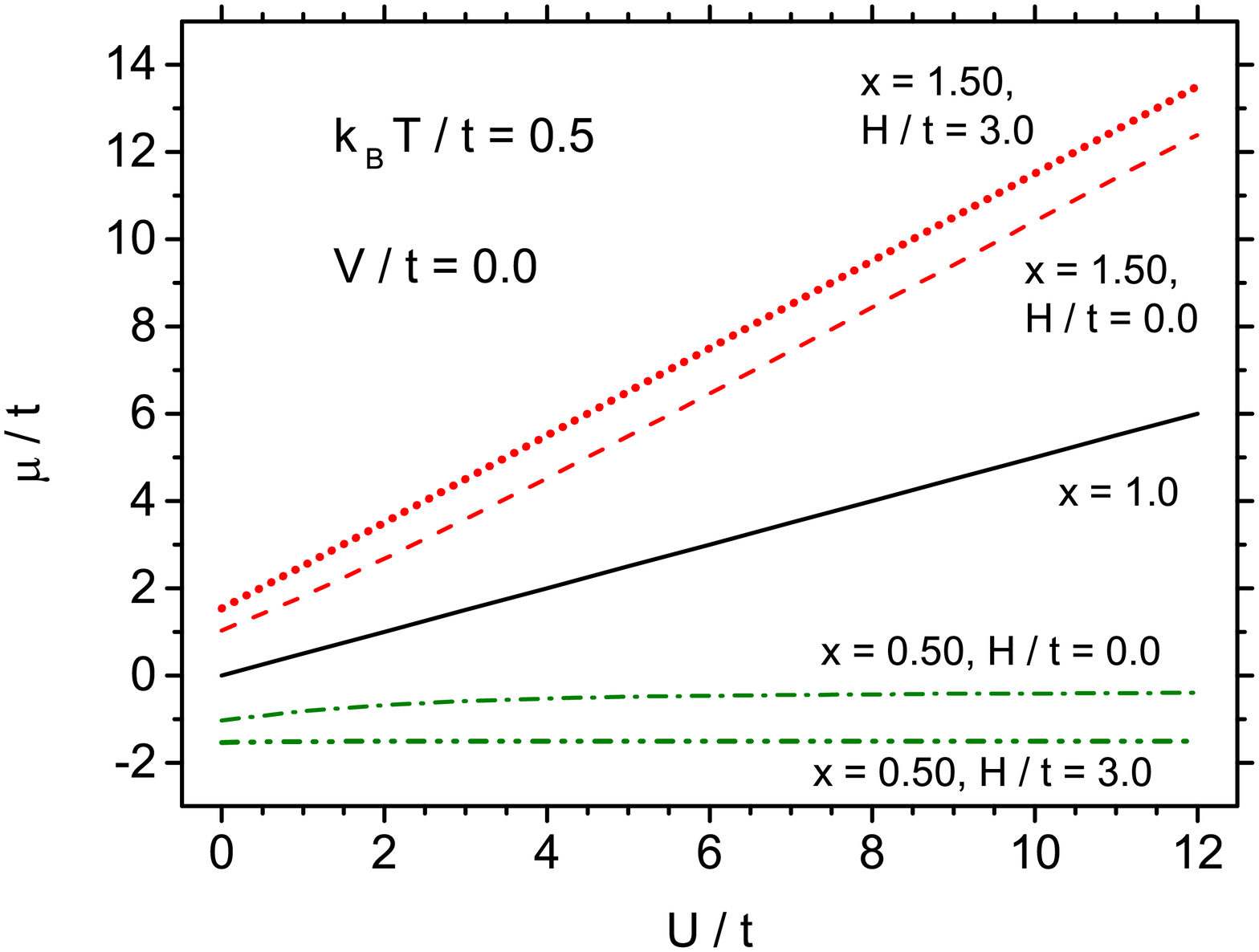}
  \end{center}
   \caption{\label{fig:fig9} Chemical potential $\mu/t$ vs. Hubbard $U/t$-parameter, for $k_{\rm B}T/t=0.5$ and different electron concentrations $x$ per atom. Two magnetic fields are chosen: $H/t=0$ and $H/t=3$, whereas the electric field is set to zero.}
\end{figure}

In Fig.~\ref{fig:fig9} the chemical potential $\mu/t$ is plotted as a function of $U/t$, for different concentrations $x$ and constant temperature $k_{\rm B}T/t$=0.5. The absence or presence of the magnetic field with rather high value, $H/t=3$, is taken into account, whereas the electric field is set to $E=0$. Only for $x=1$ the ideal linear behaviour is observed, which is in accordance with $\mu =U/2$. In general, $\mu$ is an increasing function of $U$, however, for small $x$ the chemical potential becomes only weakly dependent on $U$, which is in accordance with Fig.~\ref{fig:fig6}. The magnetic field makes the curves more flat for $x <1$ and more steep for $x >1$, with no influence observed for $x =1$. Again, the curves are symmetric with respect to the case of $x=1$. An increase of the Hubbard energy $U$, which stands for on-site Coulomb repulsion, means that also increasing chemical potential is needed for the electron to be adsorbed by the open system. A significant increase of $\mu$ is observed when 
both $U$ and $x$ are large.\\

\begin{figure}[h!]
  \begin{center}
   \includegraphics[scale=0.45]{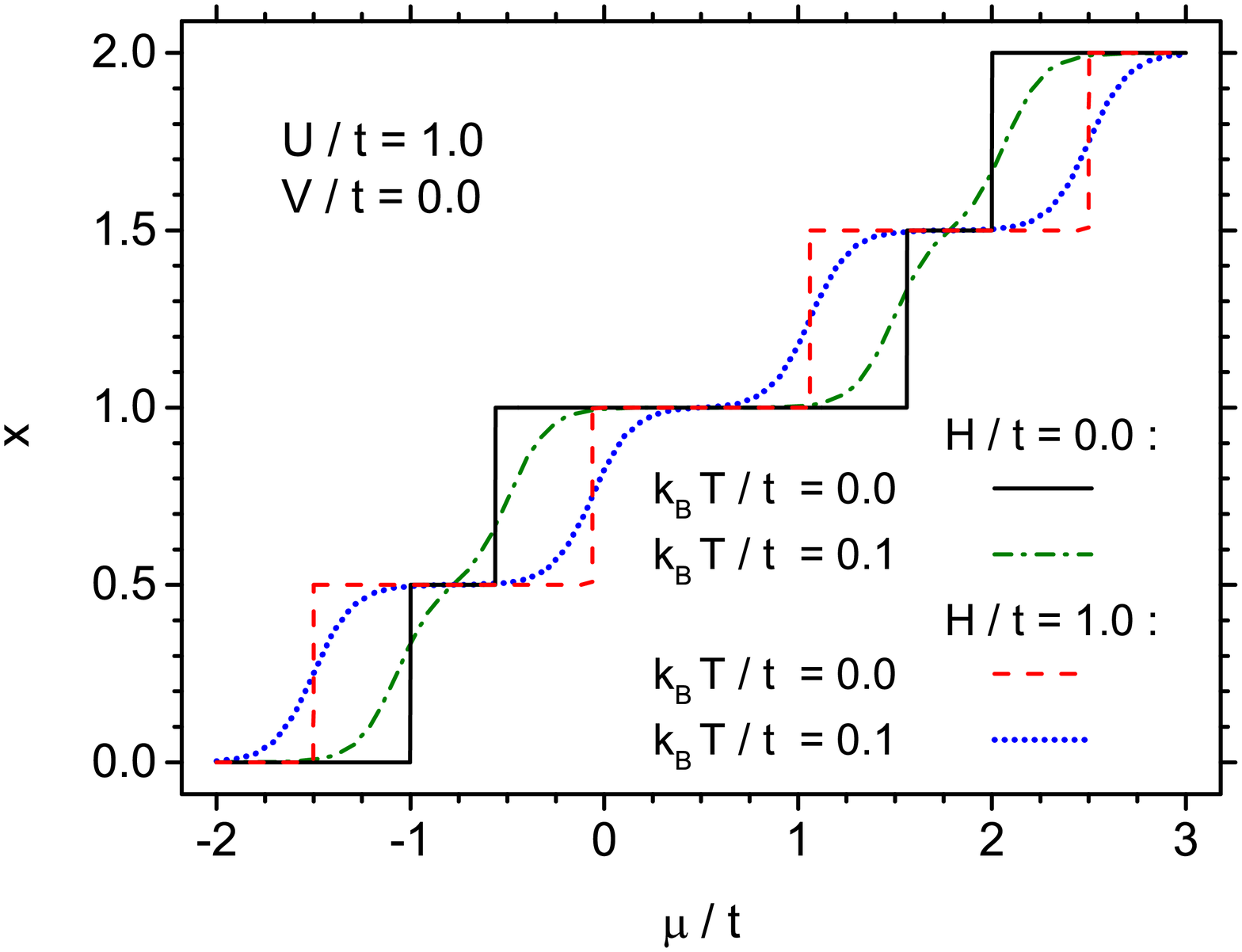}
  \end{center}
   \caption{\label{fig:fig10} A step-wise behaviour of electron concentration $x$ per atom vs. chemical potential $\mu/t$, for $U/t=1$. Different curves correspond to two chosen temperatures: $k_{\rm B}T/t=0$ and $k_{\rm B}T/t=1$, as well as two magnetic fields: $H/t=0$ and $H/t=1$. The electric field is set to zero.}
\end{figure}

Treating chemical potential as a variable, the resulting electron concentration $x$ is plotted in Fig.~\ref{fig:fig10}. In such approach the chemical potential can be understood as an uniform external potential acting on the pair of atoms, whereas the electrons can be freely exchanged between the pair-cluster and its environment. In order to preserve this condition, the electric field is set to $E=0$. Resulting mean number of electrons per atom (i.e., concentration $x$) shows  a step-wise behaviour at low temperatures, seen already at $k_{\rm B}T/t$=0.1, and, in particular, in the ground state, when $k_{\rm B}T/t$=0. After applying the external magnetic field with the value  $H/t=1$, the ranges of plateaus (horizontal parts of concentration $x$, representing the intervals of charge stability) are changed in comparison with  the case of $H=0$.  This is an effect of energy competition between the Coulombic repulsion for on-site electrons and the magnetic Zeeman term. It should be noticed that 
the external magnetic field is not able to change the height of the steps, which occur for 
$x$=0, 0.5, 1, 1.5, and 2, i.e., for the same values as in Fig.~\ref{fig:fig8}. One should also notice that the energy $U$ in Fig.~\ref{fig:fig10}, $U/t$=1, differs from that in Fig.~\ref{fig:fig8}. As it could be expected, the common point for all curves presented in Fig.~\ref{fig:fig10} lies at ($\mu/t=0.5, \, x=1$).\\

\begin{figure}[h!]
  \begin{center}
   \includegraphics[scale=0.45]{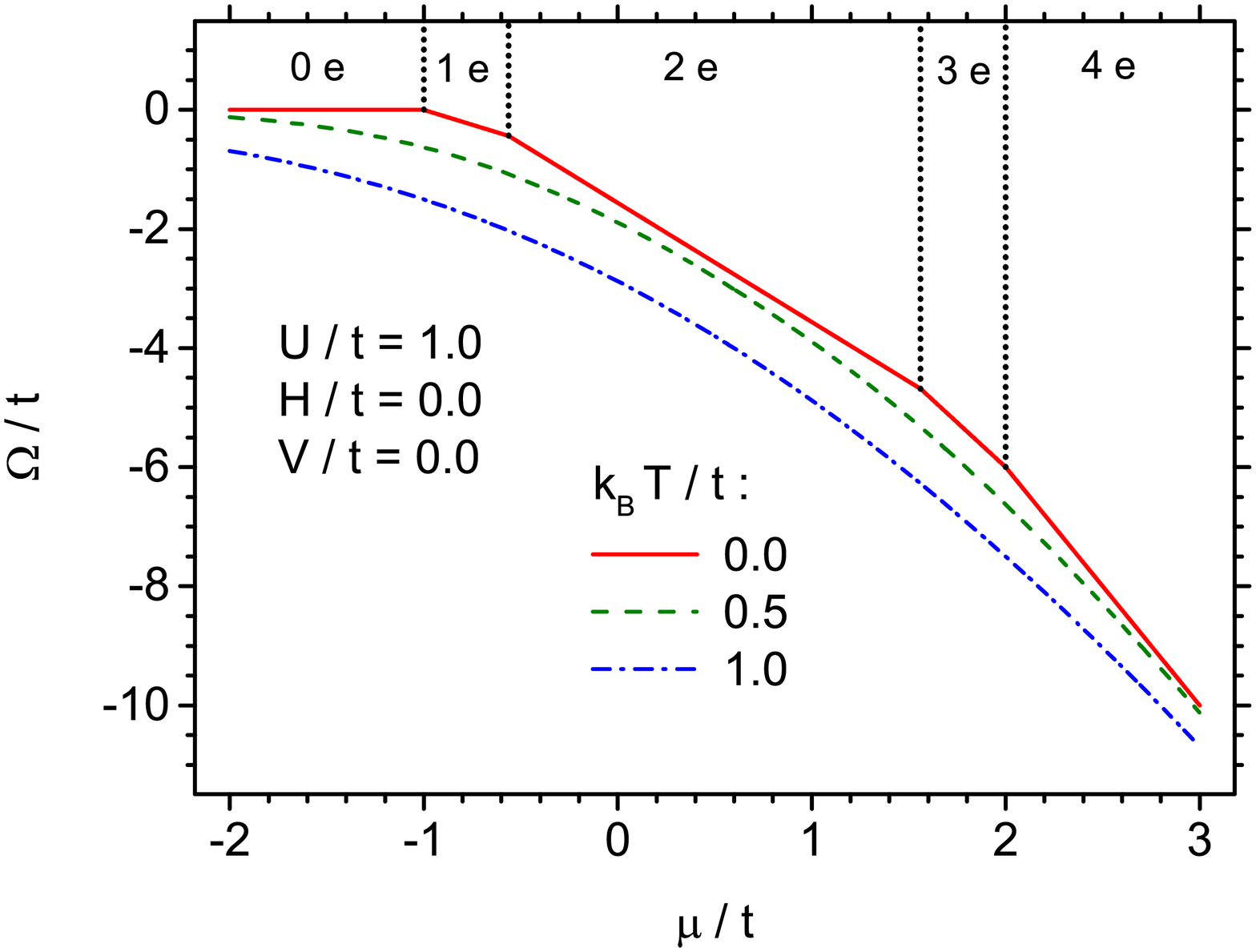}
  \end{center}
   \caption{\label{fig:fig11} Grand potential $\Omega/t$ as a function of the chemical potential $\mu/t$, for $U/t=1$ and three different temperatures. The magnetic and electric fields are set to zero.}
\end{figure}

In the next figure (Fig.~\ref{fig:fig11}) the grand potential, $\Omega/t$, is presented vs. chemical potential $\mu/t$, in the absence of  external fields but for different temperatures. The Hubbard energy $U$ amounts to $U/t=1$, thus, in the ground state, the result can be related to the corresponding $x$-curve in Fig.~\ref{fig:fig10}. It can be seen that for  $k_{\rm B}T/t$=0 the grand potential has the kinks at $\mu/t$ =-1, -0.56, 1.56, and 2, i.e., at the same discontinuity points as $x$ in Fig.~\ref{fig:fig10}. We can identify these discontinuity points as the limits of the  intervals where the integer number of electrons is present in the Hubbard pair,  whereas the system is in the ground state. Within these intervals the grand potential is presented by the straight lines, and its derivative, i.e., the number of electrons (see Eq.~\ref{eq5}), is constant. By increasing the temperature, the grand potential transforms into the smooth curve, which monotonically decreases with increase of $\mu$. We also see 
that for constant $\mu$, the increase of temperature involves always a decrease  of $\Omega$. Hence, it can be concluded that the entropy, $S=-(\partial \Omega/\partial T)_{\mu}$, will be positive for all $\mu$ and $T$.\\

\begin{figure}[h!]
  \begin{center}
   \includegraphics[scale=0.45]{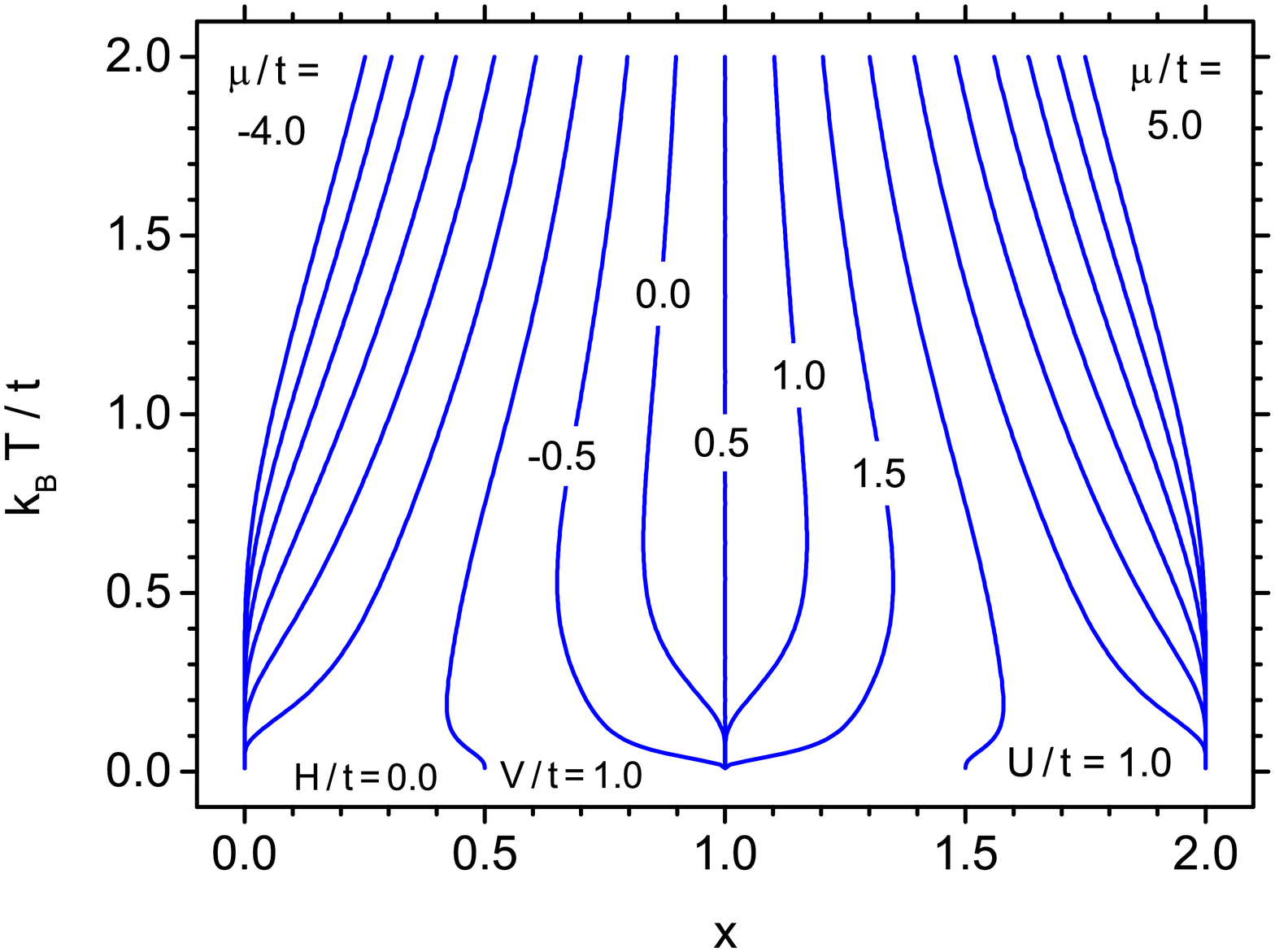}
  \end{center}
   \caption{\label{fig:fig12} Isolines presenting the constant chemical potentials $\mu/t$, in the coordinates $x$ (electron concentration per atom) and   $k_{\rm B}T/t$ (temperature). The curves are separated by $|\Delta \mu|/t=0.5$. The magnetic field is set to zero, whereas the electric field potential is $V/t=1$ and $U/t=1$.}
\end{figure}

Quantum behaviour of the electron concentration for low temperatures is also illustrated in Fig.~\ref{fig:fig12}. In this diagram the isolines representing constant chemical potentials, $\mu/t$, are plotted in the coordinates $x$ and $k_{\rm B}T/t$. The isolines are separated by $|\Delta \mu|/t=0.5$. The external  magnetic field is set to zero, whereas $V/t=1$ and the energy $U$ is $U/t=1$. It can be seen that for $T=0$ only five values of electron concentration are allowed, which are distributed symmetrically around $x=1$, and the  $\mu$ values can be divided into five groups. These groups correspond to five regions of constant electron concentrations, as discussed in Fig.~\ref{fig:fig11} for the ground state.
By increasing temperature, the discrete values of $x$ are transformed into the continuous spectrum extending over arbitrary $\mu$ value. Similarly to the ground state, the spectrum remains symmetric around ($x=1$ , $\mu/t=0.5$) for all the temperatures.\\

\begin{figure}[h!]
  \begin{center}
   \includegraphics[scale=0.45]{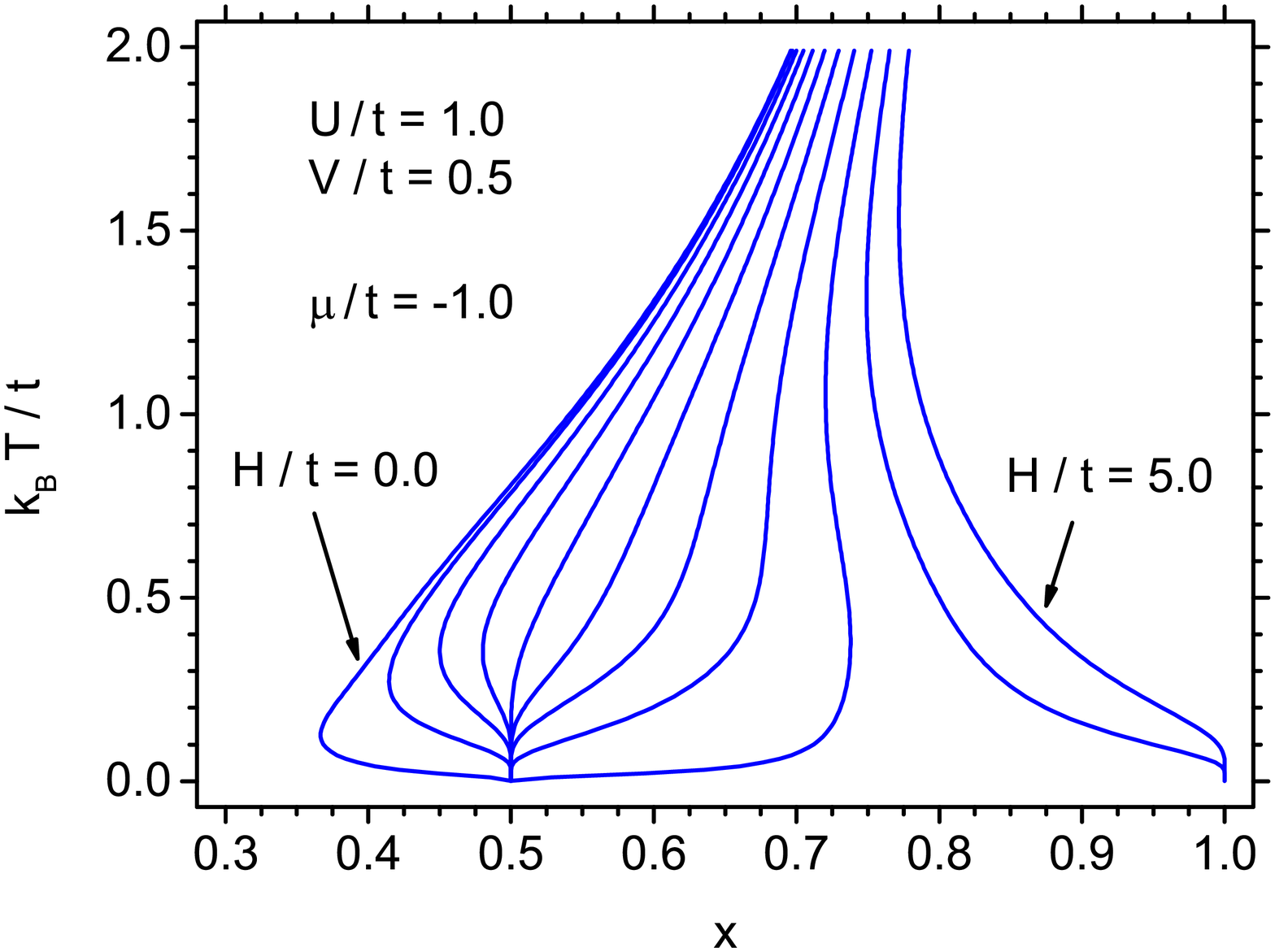}
  \end{center}
   \caption{\label{fig:fig13} Isolines presenting constant magnetic fields $H/t$, in the coordinates $x$ (electron concentration per atom) and   $k_{\rm B}T/t$ (temperature). The curves are separated by $|\Delta H|/t=0.5$. All curves correspond to the constant chemical potential $\mu/t=-1$  and $U/t=1$. The electric field potential is set to $V/t$=0.5.}
\end{figure}

In Fig.~\ref{fig:fig13}, the isolines representing the constant magnetic fields, $H/t$, which all correspond to the constant chemical potential with the value $\mu/t=-1$, are plotted in the coordinates $x$ and $k_{\rm B}T/t$. The electric field is assumed  as $V/t$=0.5 and the energy $U$ is set to $U/t=1$. Again, quantum behaviour of the electron concentration $x$ in the ground state is observed. The curves in Fig.~\ref{fig:fig13} presents an evolution of the constant chemical potential curve (with the value $\mu/t=-1$) under the influence of the magnetic field. In particular, it is demonstrated that for high temperatures the increase of the magnetic field causes a continuous increase of the electron concentration $x$. However, for small $T$, the process of increasing $x$ becomes quantized, and the highest electron concentration, which 
can be reached when $H$ increases, is $x=1$. It can also be noticed that each curve presented in previous Fig.~\ref{fig:fig12}, and described by the constant parameters $\mu$, $U$  and  $H=0$, can be studied in the external magnetic field analogously to the present case (Fig.~\ref{fig:fig13}).\\

\begin{figure}[h!]
  \begin{center}
   \includegraphics[scale=0.45]{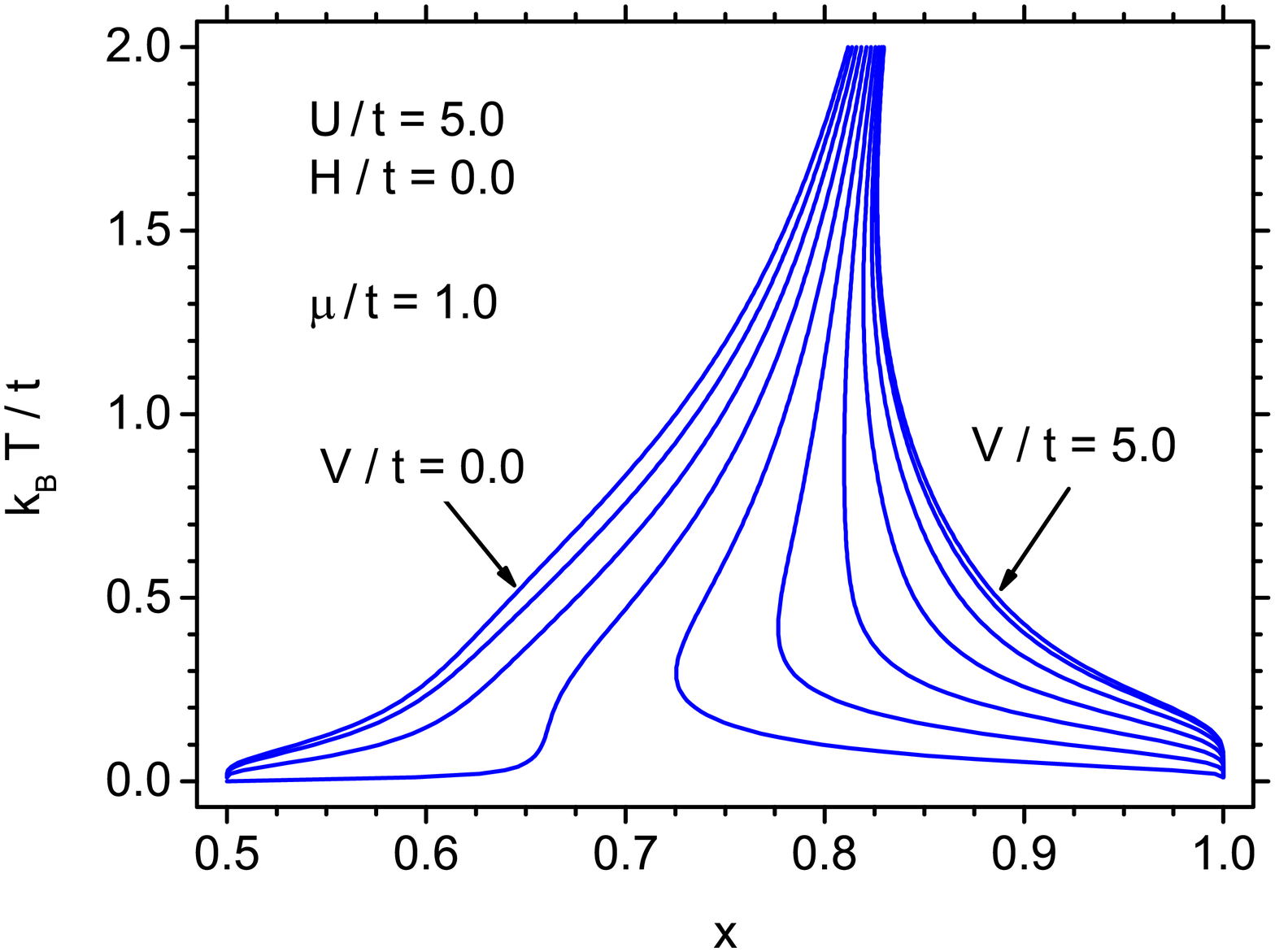}
  \end{center}
   \caption{\label{fig:fig14} Isolines presenting constant electric field potentials $V/t$, in the coordinates $x$ (electron concentration per atom) and   $k_{\rm B}T/t$ (temperature). The curves are separated by $|\Delta V|/t=0.5$. All curves correspond to the constant chemical potential $\mu/t=1$  and parameter $U/t=5$. The magnetic field is set to zero.}
\end{figure}

In the last figure (Fig.~\ref{fig:fig14}) the isolines of constant electric field potential are presented, in the coordinates $x$ and $k_{\rm B}T/t$. The isolines correspond to $U/t$=5 and constant chemical potential $\mu/t=1$, whereas the magnetic field is set to zero. It is seen that these isolines can be classified into two groups, due to quantization of electron concentration for $T=0$. Namely, for such a set of parameters, only one or two electrons can exist in the ground state (i.e., $2x$=1 or $2x$=2). The same quantization was observed in the previous figure (Fig.~\ref{fig:fig13}) Again, for higher temperatures the spectrum becomes continuous, and, at the same time, the range of allowed $x$ is narrowing.\\

\section{Summary and conclusion}

In the paper we presented some selected results of exact calculations of the chemical potential for the Hubbard pair-cluster embedded in the external fields. The Hubbard pair is treated as an open system, where the mean number of electrons results from thermodynamic equilibrium. At first, we found that the dependence of $\mu$ on the small electric field, when the external potential is applied asymmetrically to both atoms, is weaker than the dependence on the magnetic field (Fig.~\ref{fig:fig1}). However, in a wider range of $V/t$ the the influence of electric field is remarkable (Figs.~\ref{fig:fig3}, ~\ref{fig:fig7}, ~\ref{fig:fig14}).
The influence of Hubbard $U$ repulsion parameter has also been taken into account. In particular, for low temperatures the step-wise changes of the electron concentration (or, equivalently, chemical potential) are discussed.\\

It should be noticed that the curves representing the chemical potential are symmetric with respect to the half-filling case ($x$=1, $\mu=U/2$) due to a particle-hole symmetry. In particular, the consequences of this symmetry is seen in Figs.~\ref{fig:fig1}, ~\ref{fig:fig4}, ~\ref{fig:fig6} -~\ref{fig:fig10} and ~\ref{fig:fig12}.

The calculations of the chemical potential, which may be performed over a wide range of parameters simultaneously taking into account the electric and magnetic fields, are of primary importance for further statistical-thermodynamic studies of the system in question. According to the theoretical background given in this paper, further investigations of open systems will be aimed at the exact calculations of all thermodynamic properties which can be obtained from the knowledge of the grand partition function. For instance, various correlation functions, mean energy, as well as all magnetic and electric properties can be calculated (in preparation). In particular, the thermodynamic response functions, such as the magnetic or electric susceptibilities, can be studied within the method. Such investigations should broaden our knowledge about the physics of small clusters.

\appendix
\section{Construction of annihilation and creation operators}

On the basis of Pauli matrices:
\begin{equation}
\sigma_x=\left(\begin{array}{cc}0 & 1 \\ 1 & 0 \end{array}\right), \;\; 
\sigma_y=\left(\begin{array}{cc}0 & -i \\ i & 0 \end{array}\right), \;\; 
\sigma_z=\left(\begin{array}{cc}1 & 0 \\ 0 & -1 \end{array}\right), \;{\rm and}\;\; 
I=\left(\begin{array}{cc}1 & 0 \\ 0 & 1 \end{array}\right)
\label{a1}
\end{equation}
the spin deviation operators are presented as follows: 
\begin{equation}
\alpha=\frac{1}{2}\left(\sigma_x+i\sigma_y\right)=\left(\begin{array}{cc}0 & 1 \\ 0 & 0 \end{array}\right), \;\; 
\alpha^+=\frac{1}{2}\left(\sigma_x-i\sigma_y\right)=\left(\begin{array}{cc}0 & 0 \\ 1 & 0 \end{array}\right), \; {\rm and}\;\; \beta=\sigma_z.
\label{a2}
\end{equation}
We define single-particle auxiliary operators for the electron which can occupy 4 possible states ($\gamma=a,b;\; \sigma=\uparrow, \downarrow$) \cite{Suffczynski}:  
\begin{eqnarray}
\alpha_1=\alpha \otimes I \otimes I \otimes I, \;\;\;\;\; \beta_1=\beta \otimes I \otimes I \otimes I\nonumber\\
\alpha_2=I \otimes \alpha \otimes I \otimes I, \;\;\;\;\; \beta_2=I \otimes \beta \otimes I \otimes I\nonumber\\
\alpha_3=I \otimes I \otimes \alpha \otimes I, \;\;\;\;\; \beta_3=I \otimes I \otimes \beta \otimes I\nonumber\\
\alpha_4=I \otimes I \otimes I \otimes \alpha, \;\;\;\;\; \beta_4=I \otimes I \otimes I \otimes \beta,
\label{a3}
\end{eqnarray}
where $\otimes$ stands for the matrix outer product. Then, the annihilation operators in these 4 states are obtained by the ordinary matrix multiplication of the auxiliary operators:
\begin{eqnarray}
c_1=\alpha_1 \nonumber\\
c_2=\alpha_2 \beta_1 \nonumber\\
c_3=\alpha_3 \beta_1 \beta_2 \nonumber\\
c_4=\alpha_4 \beta_1 \beta_2 \beta_3. 
\label{a4}
\end{eqnarray}
The numbering of states can be chosen as follows:
\begin{equation}
c_{a,\downarrow}=c_1, \;\;\; c_{a,\uparrow}=c_2, \;\;\; c_{b,\downarrow}=c_3, \;\;\; c_{b,\uparrow}=c_4.
\label{a5}
\end{equation} 
The operators $c_{a,\downarrow}$, $c_{a,\uparrow}$, $c_{b,\downarrow}$ and $c_{b,\uparrow}$ are represented by 16 $\times$ 16 sparse matrices with real elements. The only non-zero elements of these matrices are the following:\\
$c_{a,\downarrow}$:
\begin {eqnarray}
c_{a,\downarrow}\left[9,1\right]=c_{a,\downarrow}\left[10,2\right]=c_{a,\downarrow}\left[11,3\right]=c_{a,\downarrow}\left[12,4\right]=\nonumber\\
=c_{a,\downarrow}\left[13,5\right]=c_{a,\downarrow}\left[14,6\right]=c_{a,\downarrow}\left[15,7\right]=c_{a,\downarrow}\left[16,8\right]=1;
\label{a6}
\end {eqnarray}    

$c_{a,\uparrow}$:
\begin {eqnarray}
c_{a,\uparrow}\left[5,1\right]=c_{a,\uparrow}\left[6,2\right]=c_{a,\uparrow}\left[7,3\right]=c_{a,\uparrow}\left[8,4\right]=1,\nonumber\\
c_{a,\uparrow}\left[13,9\right]=c_{a,\uparrow}\left[14,10\right]=c_{a,\uparrow}\left[15,11\right]=c_{a,\uparrow}\left[16,12\right]=-1;
\label{a7}
\end {eqnarray}    

$c_{b,\downarrow}$:
\begin {eqnarray}
c_{b,\downarrow}\left[3,1\right]=c_{b,\downarrow}\left[4,2\right]=c_{b,\downarrow}\left[15,13\right]=c_{b,\downarrow}\left[16,14\right]=1,\nonumber\\
c_{b,\downarrow}\left[7,5\right]=c_{b,\downarrow}\left[8,6\right]=c_{b,\downarrow}\left[11,9\right]=c_{b,\downarrow}\left[12,10\right]=-1;
\label{a8}
\end {eqnarray}  

$c_{b,\uparrow}$:
\begin {eqnarray}
c_{b,\uparrow}\left[2,1\right]=c_{b,\uparrow}\left[8,7\right]=c_{b,\uparrow}\left[12,11\right]=c_{b,\uparrow}\left[14,13\right]=1,\nonumber\\
c_{b,\uparrow}\left[4,3\right]=c_{b,\uparrow}\left[6,5\right]=c_{b,\uparrow}\left[10,9\right]=c_{b,\uparrow}\left[16,15\right]=-1.
\label{a9}
\end {eqnarray}  
The creation operators $c_{\gamma, \sigma}^+$ are the hermitian adjoints of $c_{\gamma, \sigma}$ and they can be obtained by the transposition of annihilation matrices \ref{a5}. It can be checked that the above matrices satisfy the anticommutation relations for fermionic operators:
\begin {equation}
\left[c_{\gamma,\sigma},c_{\gamma^{\prime},\sigma^{\prime}}\right]_+=\left[c_{\gamma,\sigma}^+,c_{\gamma^{\prime},\sigma^{\prime}}^+\right]_+=0,
\label{a10}
\end {equation}
and
\begin {equation}
\left[c_{\gamma,\sigma},c_{\gamma^{\prime},\sigma^{\prime}}^+\right]=\delta_{\gamma,\gamma^{\prime}}\delta_{\sigma,\sigma^{\prime}},
\label{a11}
\end {equation}
as well as the Pauli principle:
\begin {equation}
\left(c_{\gamma,\sigma}^+\right)^2=\left(c_{\gamma,\sigma}\right)^2=0.
\label{a12}
\end {equation}

\section{The pair Hamiltonian and its diagonalization}

Having $c_{\gamma,\sigma}$ and $c_{\gamma,\sigma}^+$ in their explicit form, the matrices representing occupation numbers with a given spin, $n_{\gamma,\sigma}$, and $z$-component of resulting spin on the $\gamma=a,b$ lattice site, $S_{\gamma}^z$,  can be easily found from their definitions:
\begin {equation}
n_{\gamma,\sigma}=c_{\gamma,\sigma}^+c_{\gamma,\sigma},
\label{a13}
\end {equation}
and
\begin {equation}
S_{\gamma}^z=\frac{1}{2}\left(n_{\gamma,\uparrow}-n_{\gamma,\downarrow}\right).
\label{a14}
\end {equation}
The total occupation number for given lattice site is then represented by the operator:
\begin {equation}
n_{\gamma}=n_{\gamma,\uparrow}+n_{\gamma,\downarrow}.
\label{a15}
\end {equation}
Thus, all operators in the pair Hamiltonian (Eq.~\ref{eq1}) can be explicitly expressed by the matrices given above. Consequently, the pair Hamiltonian can be presented as 16$\times$16 sparse matrix. However, in  open system, we should deal with the extended Hamiltonian, containing the chemical potential term, which, after division by the hopping integral, can be presented in the dimensionless form:
\begin {equation}
\widetilde {\mathcal{H}}_{a,b}=\frac{1}{t}\left[\mathcal{H}_{a,b}-\mu \left(n_a+n_b\right)\right]
\label{a16}
\end {equation}
As a result, the extended pair Hamiltonian containing interactions with the external fields has the matrix form:
 \begin {equation}
\widetilde {\mathcal{H}}_{a,b}=\left(\begin{array}{cccccccccccccccc}
d_1 & 0 & 0 & 0 & 0 & 0 & 0 & 0 & 0 & 0 & 0 & 0 & 0 & 0 & 0 & 0 \\ 
0 & d_2 & 0 & 0 & -1 & 0 & 0 & 0 & 0 & 0 & 0 & 0 & 0 & 0 & 0 & 0 \\
0 & 0 & d_3 & 0 & 0 & 0 & 0 & 0 & -1 & 0 & 0 & 0 & 0 & 0 & 0 & 0 \\
0 & 0 & 0 & d_4 & 0 & 0 & 1 & 0 & 0 & -1 & 0 & 0 & 0 & 0 & 0 & 0 \\
0 & -1 & 0 & 0 & d_5 & 0 & 0 & 0 & 0 & 0 & 0 & 0 & 0 & 0 & 0 & 0 \\ 
0 & 0 & 0 & 0 & 0 & d_6 & 0 & 0 & 0 & 0 & 0 & 0 & 0 & 0 & 0 & 0 \\
0 & 0 & 0 & 1 & 0 & 0 & d_7 & 0 & 0 & 0 & 0 & 0 & 1 & 0 & 0 & 0 \\
0 & 0 & 0 & 0 & 0 & 0 & 0 & d_8 & 0 & 0 & 0 & 0 & 0 & 1 & 0 & 0 \\
0 & 0 & -1 & 0 & 0 & 0 & 0 & 0 & d_9 & 0 & 0 & 0 & 0 & 0 & 0 & 0 \\ 
0 & 0 & 0 & -1 & 0 & 0 & 0 & 0 & 0 & d_{10} & 0 & 0 & -1 & 0 & 0 & 0 \\
0 & 0 & 0 & 0 & 0 & 0 & 0 & 0 & 0 & 0 & d_{11} & 0 & 0 & 0 & 0 & 0 \\
0 & 0 & 0 & 0 & 0 & 0 & 0 & 0 & 0 & 0 & 0 & d_{12} & 0 & 0 & 1 & 0 \\
0 & 0 & 0 & 0 & 0 & 0 & 1 & 0 & 0 & -1 & 0 & 0 & d_{13} & 0 & 0 & 0 \\ 
0 & 0 & 0 & 0 & 0 & 0 & 0 & 1 & 0 & 0 & 0 & 0 & 0 & d_{14} & 0 & 0 \\
0 & 0 & 0 & 0 & 0 & 0 & 0 & 0 & 0 & 0 & 0 & 1 & 0 & 0 & d_{15} & 0 \\
0 & 0 & 0 & 0 & 0 & 0 & 0 & 0 & 0 & 0 & 0 & 0 & 0 & 0 & 0 & d_{16} \\
\end{array}\right),
\label{a17}
\end {equation}
where the diagonal elements $d_i$ ($i=1,\ldots,16$) are listed below:
\begin {eqnarray}
d_{1}&=&0, \nonumber\\
d_{2}&=&-\mu/t+V/t-H/\left(2t\right), \nonumber\\
d_{3}&=&-\mu/t+V/t+H/\left(2t\right), \nonumber\\
d_{4}&=&U/t-2\left(\mu/t-V/t\right), \nonumber\\
d_{5}&=&-\mu/t-V/t-H/\left(2t\right), \nonumber\\
d_{6}&=&-2\mu/t-H/t, \nonumber\\
d_{7}&=&-2\mu/t, \nonumber\\
d_{8}&=&U/t-3\mu/t+V/t-H/\left(2t\right), \nonumber\\
d_{9}&=&-\mu/t-V/t+H/\left(2t\right), \nonumber\\
d_{10}&=&-2\mu/t, \nonumber\\
d_{11}&=&-2\mu/t+H/t, \nonumber\\
d_{12}&=&U/t-3\mu/t+V/t+H/\left(2t\right), \nonumber\\
d_{13}&=&U/t-2\left(\mu/t+V/t\right), \nonumber\\
d_{14}&=&U/t-3\mu/t-V/t-H/\left(2t\right), \nonumber\\
d_{15}&=&U/t-3\mu/t-V/t+H/\left(2t\right), \nonumber\\
d_{16}&=&2U/t-4\mu/t.
\label{a18}
\end {eqnarray}

The extended pair Hamiltonian $\widetilde {\mathcal{H}}_{a,b}$ fulfils the eigenequation:
\begin {equation}
\widetilde {\mathcal{H}}_{a,b} \Psi_{i}=E_{i}\Psi_{i},
\label{a19}
\end {equation}
where $E_{i}$ are dimensionless energies and $\Psi_{i}$ are the eigenvectors ($i$=1,...,16). In general, $\Psi_{i}$ can be presented as a linear combination of orthonormal basis $\varphi_{j}$ ($j$=1,...,16), i.e.,
\begin {equation}
\Psi_{i}=\sum_{j=1}^{16}C_{i,j}\varphi_{j}.
\label{a20}
\end {equation}
In (\ref{a20}) $\varphi_{j}$ are the unit vectors represented by 1-columnar, 16th-row matrices, with the only non-zero element, equal to 1, located in the $j$-th row.
\begin{equation}
\varphi_{j}=\left(\begin{array}{c} \delta_{j,1} \\  \ldots \\  \delta_{j,k} \\
\ldots  \\ \delta_{j,16} \end{array}\right).
\label{a21}
\end{equation}
It follows from the matrix form of $\widetilde {\mathcal{H}}_{a,b}$ that for $i$=1,6,11, and 16 the corresponding energies are  immediately given as: 
\begin {equation}
E_{1}=d_{1}\quad ; \quad E_{6}=d_{6} \quad ; \quad E_{11}=d_{11}  \quad {\rm and} \quad E_{16}=d_{16}.
\label{a22}
\end {equation}
Simultaneously, corresponding eigenvectors $\Psi_{i}$ have the simple form:
\begin {equation}
\Psi_{1}=\varphi_{1}\quad ; \quad \Psi_{6}=\varphi_{6} \quad ; \quad \Psi_{11}=\varphi_{11}  \quad {\rm and} \quad \Psi_{16}=\varphi_{16},
\label{a23}
\end {equation}
respectively. In turn, for the energy pairs ($E_{2},E_{5}$), ($E_{3},E_{9}$), ($E_{8},E_{14}$), and ($E_{12},E_{15}$), we obtain from Eq.~(\ref{a19}) corresponding blocks of 2 coupled equations for the coefficients $C_{i,j}$, which leads to the quadratic equations for the energy. Solving these equations one obtains:
\begin {eqnarray}
E_{2}&=&\frac{1}{2}\left(d_{2}+d_{5}\right)-\frac{1}{2}\sqrt{\left(d_{2}-d_{5}\right)^2+4}\nonumber\\
E_{5}&=&\frac{1}{2}\left(d_{2}+d_{5}\right)+\frac{1}{2}\sqrt{\left(d_{2}-d_{5}\right)^2+4}\nonumber\\
E_{3}&=&\frac{1}{2}\left(d_{3}+d_{9}\right)-\frac{1}{2}\sqrt{\left(d_{3}-d_{9}\right)^2+4}\nonumber\\
E_{9}&=&\frac{1}{2}\left(d_{3}+d_{9}\right)+\frac{1}{2}\sqrt{\left(d_{3}-d_{9}\right)^2+4}\nonumber\\
E_{8}&=&\frac{1}{2}\left(d_{8}+d_{14}\right)-\frac{1}{2}\sqrt{\left(d_{8}-d_{14}\right)^2+4}\nonumber\\
E_{14}&=&\frac{1}{2}\left(d_{8}+d_{14}\right)+\frac{1}{2}\sqrt{\left(d_{8}-d_{14}\right)^2+4}\nonumber\\
E_{12}&=&\frac{1}{2}\left(d_{12}+d_{15}\right)-\frac{1}{2}\sqrt{\left(d_{12}-d_{15}\right)^2+4}\nonumber\\
E_{15}&=&\frac{1}{2}\left(d_{12}+d_{15}\right)+\frac{1}{2}\sqrt{\left(d_{12}-d_{15}\right)^2+4}.
\label{a24}
\end {eqnarray}
Corresponding eigenvectors have the form of:
\begin {eqnarray}
\Psi_{2}&=&\frac{1}{\sqrt{1+\left(d_{5}-E_{2}\right)^2}}\left[\left(d_{5}-E_{2}\right)\varphi_{2}+\varphi_{5}\right] \nonumber\\
\Psi_{5}&=&\frac{1}{\sqrt{1+\left(d_{5}-E_{5}\right)^2}}\left[\left(d_{5}-E_{5}\right)\varphi_{2}+\varphi_{5}\right] \nonumber\\
\Psi_{3}&=&\frac{1}{\sqrt{1+\left(d_{9}-E_{3}\right)^2}}\left[\left(d_{9}-E_{3}\right)\varphi_{3}+\varphi_{9}\right] \nonumber\\
\Psi_{9}&=&\frac{1}{\sqrt{1+\left(d_{9}-E_{9}\right)^2}}\left[\left(d_{9}-E_{9}\right)\varphi_{3}+\varphi_{9}\right] \nonumber\\
\Psi_{8}&=&\frac{1}{\sqrt{1+\left(d_{14}-E_{8}\right)^2}}\left[-\left(d_{14}-E_{8}\right)\varphi_{8}+\varphi_{14}\right] \nonumber\\
\Psi_{14}&=&\frac{1}{\sqrt{1+\left(d_{14}-E_{14}\right)^2}}\left[-\left(d_{14}-E_{14}\right)\varphi_{8}+\varphi_{14}\right] \nonumber\\
\Psi_{12}&=&\frac{1}{\sqrt{1+\left(d_{15}-E_{12}\right)^2}}\left[-\left(d_{15}-E_{12}\right)\varphi_{12}+\varphi_{15}\right] \nonumber\\
\Psi_{15}&=&\frac{1}{\sqrt{1+\left(d_{15}-E_{15}\right)^2}}\left[-\left(d_{15}-E_{15}\right)\varphi_{12}+\varphi_{15}\right].
\label{a24b}
\end {eqnarray}
Remaining four energies: $E_{4}$, $E_{7}$, $E_{10}$, and $E_{13}$, can be found from Eq.~(\ref{a19}) as the solutions of the block of 4 coupled equations for coefficients $C_{i,j}$. This leads to the 4th order algebraic equation for the energy:
\begin {equation}
E^4+b\,E^3+c\,E^2+d\,E+e=0,
\label{a25}
\end {equation}
where the coefficients $b,c,d$ and $e$ can be presented as:
\begin {eqnarray}
b&=&-\left(d_{4}+d_{7}+d_{10}+d_{13}\right) \nonumber\\
c&=&d_{4}d_{7}+d_{10}d_{13}+\left(d_{4}+d_{7}\right)\left(d_{10}+d_{13}\right)-4 \nonumber\\
d&=&-d_{4}d_{7}\left(d_{10}+d_{13}\right)-d_{10}d_{13}\left(d_{4}+d_{7}\right)+2\left(d_{4}+d_{7}+d_{10}+d_{13}\right) \nonumber\\
e&=&-\left(d_{4}+d_{13}\right)\left(d_{7}+d_{10}\right)+d_{4}d_{7}d_{10}d_{13}.
\label{a26}
\end {eqnarray}
The 4th-order algebraic equation in the general form of (\ref{a25}) can be solved analytically by the methods which can be found in mathematical handbooks \cite{Abramowitz}. These quite long, but standard, methods will not be repeated here. As a result one obtains 4 analytic solutions for energies: $E_{4}$,  $E_{7}$,  $E_{10}$, and  $E_{13}$, which are real and can be used for finding corresponding eigenvectors. In order to present these eigenstates $\Psi_{4}$, $\Psi_{7}$, $\Psi_{10}$, and $\Psi_{13}$, we define auxiliary functions:
\begin {eqnarray}
f_{4}\left(E_{i}\right)&=&\frac{d_{13}-E_{i}}{d_{4}-E_{i}} \nonumber\\
f_{7}\left(E_{i}\right)&=&-\frac{d_{4}+d_{13}-2E_{i}}{\left(d_{4}-E_{i}\right)\left(d_{7}-E_{i}\right)} \nonumber\\
f_{10}\left(E_{i}\right)&=&\frac{d_{4}+d_{13}-2E_{i}}{\left(d_{4}-E_{i}\right)\left(d_{10}-E_{i}\right)}
\label{a27}
\end {eqnarray}
for $i=$ 4, 7, 10 and 13, and
\begin {equation} 
F\left(E_{i}\right)=\frac{1}{\sqrt{1+f_{4}^2\left(E_{i}\right)+f_{7}^2\left(E_{i}\right)+f_{10}^2\left(E_{i}\right)}}.
\label{a28}
\end {equation}
With these functions, the eigenvectors for $i=$4, 7, 10 and 13, can be expressed  in the form of linear combination:
\begin {equation} 
\Psi_{i}=F\left(E_{i}\right)\left[f_{4}\left(E_{i}\right)\, \varphi_{4}+f_{7}\left(E_{i}\right)\, \varphi_{7}+
f_{10}\left(E_{i}\right)\, \varphi_{10}+\varphi_{13}\right],
\label{a29}
\end {equation}
where $\varphi_{i}$ are the unit vectors as defined in Eq.~(\ref{a21}). 

It can be checked that $\Psi_{i}$ for $i=$1,...,16, form the orthonormal and complete vector set. Thus, the full spectrum of energies $E_{i}$ and corresponding eigenvectors $\Psi_{i}$ have analytically been determined for the Hubbard pair treated as an open system, whose interactions with the external fields are described by the extended dimensionless Hamiltonian $\widetilde {\mathcal{H}}_{a,b}$
(Eq.~(\ref{a16})).


\begin{thebibliography}{10}
\expandafter\ifx\csname url\endcsname\relax
  \def\url#1{\texttt{#1}}\fi
\expandafter\ifx\csname urlprefix\endcsname\relax\def\urlprefix{URL }\fi
\expandafter\ifx\csname href\endcsname\relax
  \def\href#1#2{#2} \def\path#1{#1}\fi

\bibitem{Anderson}
P.~W. Anderson, New approach to the theory of superexchange interactions, Phys.
  Rev. 115 (1959) 2--13.
\newblock \href {http://dx.doi.org/10.1103/PhysRev.115.2}
  {\path{doi:10.1103/PhysRev.115.2}}.

\bibitem{Hubbard}
J.~Hubbard, Electron correlations in narrow energy bands, Proceedings of the
  Royal Society of London A: Mathematical, Physical and Engineering Sciences
  276~(1365) (1963) 238--257.
\newblock \href {http://dx.doi.org/10.1098/rspa.1963.0204}
  {\path{doi:10.1098/rspa.1963.0204}}.

\bibitem{Gutzwiller}
M.~C. Gutzwiller, Effect of correlation on the ferromagnetism of transition
  metals, Phys. Rev. Lett. 10 (1963) 159--162.
\newblock \href {http://dx.doi.org/10.1103/PhysRevLett.10.159}
  {\path{doi:10.1103/PhysRevLett.10.159}}.

\bibitem{Kanamori}
J.~Kanamori, Electron correlation and ferromagnetism of transition metals,
  Progress of Theoretical Physics 30~(3) (1963) 275--289.
\newblock \href {http://dx.doi.org/10.1143/PTP.30.275}
  {\path{doi:10.1143/PTP.30.275}}.

\bibitem{Chen}
C.~C. Chen, M.-H. Huang, Field and temperature dependence of thermodynamic and
  correlation functions of {H}ubbard model, Journal of Applied Physics 50~(B3)
  (1979) 1761--1763.
\newblock \href {http://dx.doi.org/10.1063/1.327211}
  {\path{doi:10.1063/1.327211}}.

\bibitem{Ho}
W.-C. Ho, J.~H. Barry, Cluster-variation method applied in two-site
  approximation to the {H}ubbard model at high temperatures, Phys. Rev. B 20
  (1979) 2118--2128.
\newblock \href {http://dx.doi.org/10.1103/PhysRevB.20.2118}
  {\path{doi:10.1103/PhysRevB.20.2118}}.

\bibitem{Robaszkiewicz1}
S.~Robaszkiewicz, R.~Micnas, K.~A. Chao, Thermodynamic properties of the
  extended {H}ubbard model with strong intra-atomic attraction and an arbitrary
  electron density, Phys. Rev. B 23 (1981) 1447--1458.
\newblock \href {http://dx.doi.org/10.1103/PhysRevB.23.1447}
  {\path{doi:10.1103/PhysRevB.23.1447}}.

\bibitem{Robaszkiewicz2}
S.~Robaszkiewicz, R.~Micnas, K.~A. Chao, Chemical potential and order parameter
  of extended {H}ubbard model with strong intra-atomic attraction, Phys. Rev. B
  24 (1981) 1579--1582.
\newblock \href {http://dx.doi.org/10.1103/PhysRevB.24.1579}
  {\path{doi:10.1103/PhysRevB.24.1579}}.

\bibitem{Hirsch3}
J.~E. Hirsch, Renormalization-group study of the {H}ubbard model, Phys. Rev. B
  22 (1980) 5259--5266.
\newblock \href {http://dx.doi.org/10.1103/PhysRevB.22.5259}
  {\path{doi:10.1103/PhysRevB.22.5259}}.

\bibitem{Hirsch}
J.~E. Hirsch, Two-dimensional {H}ubbard model: Numerical simulation study,
  Phys. Rev. B 31 (1985) 4403--4419.
\newblock \href {http://dx.doi.org/10.1103/PhysRevB.31.4403}
  {\path{doi:10.1103/PhysRevB.31.4403}}.

\bibitem{Hirsch2}
J.~E. Hirsch, S.~Tang, Antiferromagnetism in the two-dimensional {H}ubbard
  model, Phys. Rev. Lett. 62 (1989) 591--594.
\newblock \href {http://dx.doi.org/10.1103/PhysRevLett.62.591}
  {\path{doi:10.1103/PhysRevLett.62.591}}.

\bibitem{Lieb}
E.~H. Lieb, Two theorems on the {H}ubbard model, Phys. Rev. Lett. 62 (1989)
  1201--1204.
\newblock \href {http://dx.doi.org/10.1103/PhysRevLett.62.1201}
  {\path{doi:10.1103/PhysRevLett.62.1201}}.

\bibitem{Lieb2}
E.~H. Lieb, F.~Y. Wu, Absence of {M}ott transition in an exact solution of the
  short-range, one-band model in one dimension, Phys. Rev. Lett. 20 (1968)
  1445--1448.
\newblock \href {http://dx.doi.org/10.1103/PhysRevLett.20.1445}
  {\path{doi:10.1103/PhysRevLett.20.1445}}.

\bibitem{Lieb3}
E.~H. Lieb, F.~Wu, The one-dimensional {H}ubbard model: a reminiscence, Physica
  A: Statistical Mechanics and its Applications 321~(1–2) (2003) 1 -- 27.
\newblock \href {http://dx.doi.org/10.1016/S0378-4371(02)01785-5}
  {\path{doi:10.1016/S0378-4371(02)01785-5}}.

\bibitem{Sorella}
S.~Sorella, E.~Tosatti, Semi-metal-insulator transition of the {H}ubbard model
  in the honeycomb lattice, EPL (Europhysics Letters) 19~(8) (1992) 699.
\newblock \href {http://dx.doi.org/10.1209/0295-5075/19/8/007}
  {\path{doi:10.1209/0295-5075/19/8/007}}.

\bibitem{Pelizzola}
A.~Pelizzola, The half-filled {H}ubbard model in the pair approximation of the
  cluster variation method, Journal of Physics A: Mathematical and General
  26~(9) (1993) 2061.
\newblock \href {http://dx.doi.org/10.1088/0305-4470/26/9/005}
  {\path{doi:10.1088/0305-4470/26/9/005}}.

\bibitem{Janis}
V.~Jani{\v{s}}, D.~Vollhardt, Construction of analytically tractable mean-field
  theories for quantum models, Zeitschrift f{\"u}r Physik B Condensed Matter
  91~(3) (1993) 317--323.
\newblock \href {http://dx.doi.org/10.1007/BF01344060}
  {\path{doi:10.1007/BF01344060}}.

\bibitem{Staudt}
{Staudt, R.}, {Dzierzawa, M.}, {Muramatsu, A.}, Phase diagram of the
  three-dimensional {H}ubbard model at half filling, Eur. Phys. J. B 17~(3)
  (2000) 411--415.
\newblock \href {http://dx.doi.org/10.1007/s100510070120}
  {\path{doi:10.1007/s100510070120}}.

\bibitem{Peres}
N.~M.~R. Peres, M.~A.~N. Ara\'ujo, D.~Bozi, Phase diagram and magnetic
  collective excitations of the {H}ubbard model for graphene sheets and layers,
  Phys. Rev. B 70 (2004) 195122.
\newblock \href {http://dx.doi.org/10.1103/PhysRevB.70.195122}
  {\path{doi:10.1103/PhysRevB.70.195122}}.

\bibitem{Kent}
P.~R.~C. Kent, M.~Jarrell, T.~A. Maier, T.~Pruschke, Efficient calculation of
  the antiferromagnetic phase diagram of the three-dimensional {H}ubbard model,
  Phys. Rev. B 72 (2005) 060411.
\newblock \href {http://dx.doi.org/10.1103/PhysRevB.72.060411}
  {\path{doi:10.1103/PhysRevB.72.060411}}.

\bibitem{Zaleski}
T.~A. Zaleski, T.~K. Kope\'c, N\'eel order in the {H}ubbard model within a
  spin-charge rotating reference frame approach: Crossover from weak to strong
  coupling, Phys. Rev. B 77 (2008) 125120.
\newblock \href {http://dx.doi.org/10.1103/PhysRevB.77.125120}
  {\path{doi:10.1103/PhysRevB.77.125120}}.

\bibitem{Schumann}
R.~Schumann, D.~Zwicker, The {H}ubbard model extended by nearest-neighbor
  {C}oulomb and exchange interaction on a cubic cluster – rigorous and exact
  results, Annalen der Physik 522~(6) (2010) 419--439.
\newblock \href {http://dx.doi.org/10.1002/andp.201010452}
  {\path{doi:10.1002/andp.201010452}}.

\bibitem{Schumann2}
R.~Schumann, Rigorous solution of a {H}ubbard model extended by
  nearest-neighbour {C}oulomb and exchange interaction on a triangle and
  tetrahedron, Annalen der Physik 17~(4) (2008) 221--259.
\newblock \href {http://dx.doi.org/10.1002/andp.200710281}
  {\path{doi:10.1002/andp.200710281}}.

\bibitem{Cisarova}
J.~\v{C}is\'{a}rov\'{a}, J.~Stre\v{c}ka, Exact solution of a coupled
  spin–electron linear chain composed of localized {I}sing spins and mobile
  electrons, Physics Letters A 378~(38–39) (2014) 2801 -- 2807.
\newblock \href {http://dx.doi.org/10.1016/j.physleta.2014.07.049}
  {\path{doi:10.1016/j.physleta.2014.07.049}}.

\bibitem{Cencarikova}
H.~\v{C}en\v{c}arikov\'{a}, J.~Stre\v{c}ka, M.~L. Lyra, Reentrant phase
  transitions of a coupled spin-electron model on doubly decorated planar
  lattices with two or three consecutive critical points, Journal of Magnetism
  and Magnetic Materials 401 (2016) 1106 -- 1122.
\newblock \href {http://dx.doi.org/10.1016/j.jmmm.2015.11.018}
  {\path{doi:10.1016/j.jmmm.2015.11.018}}.

\bibitem{Galisova}
L.~G\'{a}lisov\'{a}, J.~Stre\v{c}ka, Magnetic gr\"uneisen parameter and
  magnetocaloric properties of a coupled spin–electron double-tetrahedral
  chain, Physics Letters A 379~(39) (2015) 2474 -- 2478.
\newblock \href {http://dx.doi.org/10.1016/j.physleta.2015.07.007}
  {\path{doi:10.1016/j.physleta.2015.07.007}}.

\bibitem{Galisova2}
L.~G\'alisov\'a, J.~Stre\v{c}ka, Vigorous thermal excitations in a
  double-tetrahedral chain of localized {I}sing spins and mobile electrons
  mimic a temperature-driven first-order phase transition, Phys. Rev. E 91
  (2015) 022134.
\newblock \href {http://dx.doi.org/10.1103/PhysRevE.91.022134}
  {\path{doi:10.1103/PhysRevE.91.022134}}.

\bibitem{Harris}
A.~B. Harris, R.~V. Lange, Single-particle excitations in narrow energy bands,
  Phys. Rev. 157 (1967) 295--314.
\newblock \href {http://dx.doi.org/10.1103/PhysRev.157.295}
  {\path{doi:10.1103/PhysRev.157.295}}.

\bibitem{Silantev}
A.~V. Silant'ev, A dimer in the extended {H}ubbard model, Russian Physics
  Journal 57~(11) (2015) 1491--1502.
\newblock \href {http://dx.doi.org/10.1007/s11182-015-0406-z}
  {\path{doi:10.1007/s11182-015-0406-z}}.

\bibitem{Hasegawa}
H.~Hasegawa, Nonextensive thermodynamics of the two-site {H}ubbard model,
  Physica A: Statistical Mechanics and its Applications 351~(2–4) (2005) 273 --
  293.
\newblock \href {http://dx.doi.org/10.1016/j.physa.2005.01.025}
  {\path{doi:10.1016/j.physa.2005.01.025}}.

\bibitem{Hasegawa2}
H.~Hasegawa, Thermal entanglement of {H}ubbard dimers in the nonextensive
  statistics, Physica A: Statistical Mechanics and its Applications 390~(8)
  (2011) 1486 -- 1503.
\newblock \href {http://dx.doi.org/10.1016/j.physa.2010.12.033}
  {\path{doi:10.1016/j.physa.2010.12.033}}.

\bibitem{Spalek}
J.~Spa\l{}ek, A.~Ole\'s, K.~Chao, Thermodynamic properties of a two-site
  {H}ubbard model with orbital degeneracy, Physica A: Statistical Mechanics and
  its Applications 97~(3) (1979) 552 -- 564.
\newblock \href {http://dx.doi.org/10.1016/0378-4371(79)90095-5}
  {\path{doi:10.1016/0378-4371(79)90095-5}}.

\bibitem{Longhi}
S.~Longhi, G.~Della~Valle, V.~Foglietti, Classical realization of two-site
  {F}ermi-{H}ubbard systems, Phys. Rev. B 84 (2011) 033102.
\newblock \href {http://dx.doi.org/10.1103/PhysRevB.84.033102}
  {\path{doi:10.1103/PhysRevB.84.033102}}.

\bibitem{Juliano}
R.~Juliano, A.~de~Arruda, L.~Craco, Coexistence and competition of on-site and
  intersite {C}oulomb interactions in {M}ott-molecular-dimers, Solid State
  Communications 227 (2016) 51 -- 55.
\newblock \href {http://dx.doi.org/10.1016/j.ssc.2015.11.021}
  {\path{doi:10.1016/j.ssc.2015.11.021}}.

\bibitem{Kozlov}
M.~E. Kozlov, V.~A. Ivanov, K.~Yakushi, Development of a two-site {H}ubbard
  model for analysis of the electron-molecular vibration coupling in organic
  charge-transfer salts, Physics Letters A 214~(3) (1996) 167 -- 174.
\newblock \href {http://dx.doi.org/10.1016/0375-9601(96)00113-2}
  {\path{doi:10.1016/0375-9601(96)00113-2}}.

\bibitem{Li}
J.~Li, C.~Aron, G.~Kotliar, J.~E. Han, Electric-field-driven resistive
  switching in the dissipative {H}ubbard model, Phys. Rev. Lett. 114 (2015)
  226403.
\newblock \href {http://dx.doi.org/10.1103/PhysRevLett.114.226403}
  {\path{doi:10.1103/PhysRevLett.114.226403}}.

\bibitem{Joura}
A.~V. Joura, J.~K. Freericks, A.~I. Lichtenstein, Long-lived nonequilibrium
  states in the {H}ubbard model with an electric field, Phys. Rev. B 91 (2015)
  245153.
\newblock \href {http://dx.doi.org/10.1103/PhysRevB.91.245153}
  {\path{doi:10.1103/PhysRevB.91.245153}}.

\bibitem{Alvarez}
B.~Alvarez-Fern\'{a}ndez, J.~A. Blanco, The {H}ubbard model for the hydrogen
  molecule, European Journal of Physics 23~(1) (2002) 11.
\newblock \href {http://dx.doi.org/10.1088/0143-0807/23/1/302}
  {\path{doi:10.1088/0143-0807/23/1/302}}.

\bibitem{McKenzie}
R.~McKenzie, A strongly correlated electron model for the layered organic
  superconductors {kappa-(BEDT-TTF)2X}, Comments on Condensed Matter Physics 18
  (1998) 309.

\bibitem{Fuchs}
S.~Fuchs, E.~Gull, L.~Pollet, E.~Burovski, E.~Kozik, T.~Pruschke, M.~Troyer,
  Thermodynamics of the {3D} {H}ubbard model on approaching the {N}\'eel
  transition, Phys. Rev. Lett. 106 (2011) 030401.
\newblock \href {http://dx.doi.org/10.1103/PhysRevLett.106.030401}
  {\path{doi:10.1103/PhysRevLett.106.030401}}.

\bibitem{Rohringer}
G.~Rohringer, A.~Toschi, A.~Katanin, K.~Held, Critical properties of the
  half-filled {H}ubbard model in three dimensions, Phys. Rev. Lett. 107 (2011)
  256402.
\newblock \href {http://dx.doi.org/10.1103/PhysRevLett.107.256402}
  {\path{doi:10.1103/PhysRevLett.107.256402}}.

\bibitem{Kozik}
E.~Kozik, E.~Burovski, V.~W. Scarola, M.~Troyer, N\'eel temperature and
  thermodynamics of the half-filled three-dimensional {H}ubbard model by
  diagrammatic determinant monte carlo, Phys. Rev. B 87 (2013) 205102.
\newblock \href {http://dx.doi.org/10.1103/PhysRevB.87.205102}
  {\path{doi:10.1103/PhysRevB.87.205102}}.

\bibitem{Karchev}
N.~Karchev, Quantum critical behavior in three-dimensional one-band {H}ubbard
  model at half-filling, Annals of Physics 333 (2013) 206 -- 220.
\newblock \href {http://dx.doi.org/10.1016/j.aop.2013.03.005}
  {\path{doi:10.1016/j.aop.2013.03.005}}.

\bibitem{Yamada}
A.~Yamada, Magnetic properties and {M}ott transition in the {H}ubbard model on
  the anisotropic triangular lattice, Phys. Rev. B 89 (2014) 195108.
\newblock \href {http://dx.doi.org/10.1103/PhysRevB.89.195108}
  {\path{doi:10.1103/PhysRevB.89.195108}}.

\bibitem{Claveau}
Y.~Claveau, B.~Arnaud, S.~D. Matteo, Mean-field solution of the {H}ubbard
  model: the magnetic phase diagram, European Journal of Physics 35~(3) (2014)
  035023.
\newblock \href {http://dx.doi.org/10.1088/0143-0807/35/3/035023}
  {\path{doi:10.1088/0143-0807/35/3/035023}}.

\bibitem{Shastry}
B.~S. Shastry, Exact integrability of the one-dimensional {H}ubbard model,
  Phys. Rev. Lett. 56 (1986) 2453--2455.
\newblock \href {http://dx.doi.org/10.1103/PhysRevLett.56.2453}
  {\path{doi:10.1103/PhysRevLett.56.2453}}.

\bibitem{Su}
G.~Su, B.-H. Zhao, M.-L. Ge, Exact solution of the one-dimensional {H}ubbard
  model in a magnetic field, Phys. Rev. B 46 (1992) 14909--14911.
\newblock \href {http://dx.doi.org/10.1103/PhysRevB.46.14909}
  {\path{doi:10.1103/PhysRevB.46.14909}}.

\bibitem{Mancini}
{Mancini, F.}, {Mancini, F. P.}, Extended {H}ubbard model in the presence of a
  magnetic field, Eur. Phys. J. B 68~(3) (2009) 341--351.
\newblock \href {http://dx.doi.org/10.1140/epjb/e2008-00423-3}
  {\path{doi:10.1140/epjb/e2008-00423-3}}.

\bibitem{Tocchio}
L.~F. Tocchio, H.~Feldner, F.~Becca, R.~Valent\'{\i}, C.~Gros, Spin-liquid
  versus spiral-order phases in the anisotropic triangular lattice, Phys. Rev.
  B 87 (2013) 035143.
\newblock \href {http://dx.doi.org/10.1103/PhysRevB.87.035143}
  {\path{doi:10.1103/PhysRevB.87.035143}}.

\bibitem{Dang}
H.~T. Dang, X.~Y. Xu, K.-S. Chen, Z.~Y. Meng, S.~Wessel, Mott transition in the
  triangular lattice {H}ubbard model: A dynamical cluster approximation study,
  Phys. Rev. B 91 (2015) 155101.
\newblock \href {http://dx.doi.org/10.1103/PhysRevB.91.155101}
  {\path{doi:10.1103/PhysRevB.91.155101}}.

\bibitem{Mermin}
N.~D. Mermin, H.~Wagner, Absence of ferromagnetism or antiferromagnetism in
  one- or two-dimensional isotropic {H}eisenberg models, Phys. Rev. Lett. 17
  (1966) 1133--1136.
\newblock \href {http://dx.doi.org/10.1103/PhysRevLett.17.1133}
  {\path{doi:10.1103/PhysRevLett.17.1133}}.

\bibitem{Nolting}
A.~R. W.~Nolting, Quantum Theory of Magnetism, Springer-Verlag, Berlin, 2009.

\bibitem{Dombrowsky}
K.~Dombrowsky, S.and~Dichtel, Cumulant calculations of thermodynamic quantities
  for the {H}ubbard and the {E}mery model, Journal of Superconductivity 9~(4)
  (1996) 453--456.
\newblock \href {http://dx.doi.org/10.1007/BF00727295}
  {\path{doi:10.1007/BF00727295}}.

\bibitem{Feldner}
H.~Feldner, Z.~Y. Meng, A.~Honecker, D.~Cabra, S.~Wessel, F.~F. Assaad,
  Magnetism of finite graphene samples: Mean-field theory compared with exact
  diagonalization and quantum {M}onte {C}arlo simulations, Phys. Rev. B 81
  (2010) 115416.
\newblock \href {http://dx.doi.org/10.1103/PhysRevB.81.115416}
  {\path{doi:10.1103/PhysRevB.81.115416}}.

\bibitem{Szalowski}
K.~Sza\l{}owski, Graphene nanoflakes in external electric and magnetic in-plane
  fields, Journal of Magnetism and Magnetic Materials 382 (2015) 318 -- 327.
\newblock \href {http://dx.doi.org/10.1016/j.jmmm.2015.01.080}
  {\path{doi:10.1016/j.jmmm.2015.01.080}}.

\bibitem{Barnas}
I.~Weymann, J.~Barna\ifmmode~\acute{s}\else \'{s}\fi{}, S.~Krompiewski,
  Transport through graphenelike flakes with intrinsic spin-orbit interactions,
  Phys. Rev. B 92 (2015) 045427.
\newblock \href {http://dx.doi.org/10.1103/PhysRevB.92.045427}
  {\path{doi:10.1103/PhysRevB.92.045427}}.

\bibitem{Chao}
K.~A. Chao, J.~Spa\l{}ek, A.~M. Ole\'s, Kinetic exchange interaction in a
  narrow s-band, Journal of Physics C: Solid State Physics 10~(10) (1977) L271.
\newblock \href {http://dx.doi.org/10.1088/0022-3719/10/10/002}
  {\path{doi:10.1088/0022-3719/10/10/002}}.

\bibitem{Yosida}
K.~Yosida, Theory of Magnetism, Springer-Verlag, Berlin, 1998.

\bibitem{Tasaki}
H.~Tasaki, The {H}ubbard model - an introduction and selected rigorous results,
  Journal of Physics: Condensed Matter 10~(20) (1998) 4353.
\newblock \href {http://dx.doi.org/10.1088/0953-8984/10/20/004}
  {\path{doi:10.1088/0953-8984/10/20/004}}.

\bibitem{Micnas}
R.~Micnas, J.~Ranninger, S.~Robaszkiewicz, Superconductivity in narrow-band
  systems with local nonretarded attractive interactions, Rev. Mod. Phys. 62
  (1990) 113--171.
\newblock \href {http://dx.doi.org/10.1103/RevModPhys.62.113}
  {\path{doi:10.1103/RevModPhys.62.113}}.

\bibitem{Georges}
A.~Georges, G.~Kotliar, W.~Krauth, M.~J. Rozenberg, Dynamical mean-field theory
  of strongly correlated fermion systems and the limit of infinite dimensions,
  Rev. Mod. Phys. 68 (1996) 13--125.
\newblock \href {http://dx.doi.org/10.1103/RevModPhys.68.13}
  {\path{doi:10.1103/RevModPhys.68.13}}.

\bibitem{Hirschmeier}
D.~Hirschmeier, H.~Hafermann, E.~Gull, A.~I. Lichtenstein, A.~E. Antipov,
  Mechanisms of finite-temperature magnetism in the three-dimensional {H}ubbard
  model, Phys. Rev. B 92 (2015) 144409.
\newblock \href {http://dx.doi.org/10.1103/PhysRevB.92.144409}
  {\path{doi:10.1103/PhysRevB.92.144409}}.

\bibitem{Mielke}
A.~Mielke, The {H}ubbard model and its properties, in: P.~C. E.~Pavarini,
  E.~Koch (Ed.), Many-Body Physics: From Kondo to Hubbard, Modeling and
  Simulation, Vol.~5, Forschungszentrum J\"ulich, 2015.

\bibitem{Wolfram}
Mathematica, Version 8.0.4, Wolfram Research, Inc., Champaign, IL, 2010.

\bibitem{Suffczynski}
M.~Suffczy\'nski, Electrons in Solids (Monographs in Physics, Vol. 1),
  Ossolineum, Wroc\l{}aw, 1985.

\bibitem{Abramowitz}
Solutions of quartic equations, in: M.~Abramovitz, I.~A. Stegun (Eds.),
  Handbook of Mathematical Functions with Formulas, Graphs, and Mathematical
  Tables, New York: Dover, 1972, Ch. 3.8.3.

\end{thebibliography}
\end{document}